\documentclass[a4paper,12pt]{article}
\usepackage{geometry}

\usepackage{graphicx}
\usepackage{upgreek}

\usepackage{amssymb}
\usepackage{amsmath}
\usepackage{epsfig} 
\usepackage{epstopdf} 
\usepackage{graphicx}

\newcommand{\nw}{g}
\newcommand{\nh}{2}

\usepackage{amsmath,environ}
\NewEnviron{eqn}{\begin{equation}\begin{split} \BODY\end{split}\end{equation}}

\usepackage{color}
\usepackage{xcolor}
\usepackage{mdframed} 
\newmdtheoremenv[linecolor=white,leftmargin=-10,rightmargin=-10,backgroundcolor=white!10,innertopmargin=0pt,ntheorem]{sidenote}{Remark}[section]

\usepackage{hyperref}
\hypersetup{
colorlinks,
linkcolor={blue},
citecolor={blue},
urlcolor={blue!80!black}
}

\numberwithin{equation}{section}

\usepackage{fancyhdr}
\makeatletter
\let\runauthor\@author
\let\runtitle\@title
\makeatother
\title{A dislocation-dipole in one dimensional lattice model
\vspace{-.05in}}
\author{Basant Lal Sharma\thanks{Email: bls@iitk.ac.in}\vspace{-.05in}\\
{\small Department of Mechanical Engineering}, \vspace{-.05in}\\
{\small Indian Institute of Technology Kanpur}, \vspace{-.05in}\\
{\small Kanpur, U.P. 208016, India\vspace{-.05in}}
}
\date{}

\begin{document}

\maketitle

\begin{abstract}
A family of equilibria corresponding to dislocation-dipole, with variable separation between the two dislocations of opposite sign, is constructed in a one dimensional lattice model. A suitable path connecting certain members of this family is found which exhibits the familiar Peierls relief.
A landscape for the variation of energy has been presented to highlight certain sequential transition between these equilibria that allows an interpretation in terms of quasi-statically separating pair of dislocations of opposite sign from the viewpoint of closely related Frenkel-Kontorova model. Closed form expressions are provided for the case of a piecewise-quadratic potential wherein an analysis of the effect of an intermediate spinodal region is included.
\end{abstract}

\section*{Introduction}
The subject of defect nucleation and quasi-static propagation of defects forms the core of the subject of plasticity as well as the wider topic of irreversibility in nature. The emergence of a {\em dislocation-dipole}, referring to a configuration of two dislocations of opposite sign, is fundamental to the study of dislocation nucleation. In this context, it is relevant to recall two achievements that occurred several decades ago. First is the mechanism that Frank and Read \cite{frank7} suggested for the nucleation of a dislocation loop from an existing dislocation. Second, arguably less known \cite{trusknote}, is a classical model corresponding to a dislocation-dipole that has been presented by Nabarro \cite{nabarroFRN}, in the framework of Peierls' model of dislocation \cite{peierls,nabarro1}. Within one dimensional models, such as the Frenkel-Kontorova model \cite{FK}, an equivalent entity is a kink-antikink pair, sometimes the same is referred as dislocation-dipole too. Some numerical experiments, for example see \cite{dmitriev1,dmitriev2}, have also demonstrated the possibility of the creation of a kink-antikink pair in the Frenkel-Kontorova model due to the interaction of two breathers, but in the absence of a driving force. For a two dimensional Frenkel-Kontorova model at non-zero temperature, nucleation and propagation of kink-antikink pair has been studied through numerical simulations as well,
for example see \cite{gorno_kats_krav_tref}. It is generally argued by physical considerations that the presence of thermal fluctuations can trigger changes in the lattice configurations that lets the particles explore the energy landscape. Naturally, the space of lattice configurations also contains the metastable equilibria which are interlaced with unstable (saddle point) equilibria leading to certain barriers for changes between `neighboring' lattice equilibria and consequent `lattice trapping'. As described by Seeger and Schiller \cite{seeg_schill}, the calculation of the kink-antikink pair generation rate at a finite temperature may be related to energy barriers in the energy landscape. These conceptual foundations prepare the background and a motivation for this paper where several simple and analytically tractable results have been presented; these are anticipated to build an understanding of the effect of lattice trapping and energy barriers in this context by further augmenting the results, from a discrete viewpoint, in the corresponding continuum models \cite{Hirth}. 

Going back to the model of Frenkel and Kontorova \cite{FK}, recall that it involves an onsite potential which is periodic, in particular it assumes more than one energy well. Keeping the objective underlined above, and with an explanation in next few sentences, it is  suffices for the purpose of this paper to consider the situation when just two energy wells are permitted for each particle in lattice. 
To utilize a workable vocabulary in the paper, when two particles lie in `two different energy wells' of onsite potential, it is referred as that the corresponding particles are in `different phases'. 
It is immediately clear then that at least two types of equilibrium configurations can be studied in such a one dimensional lattice with onsite potential.
In one case the particles at $\pm\infty$ are in different phases so that this equilibrium configuration may be interpreted as dislocation or kink; in fact, it has been extensively studied during last five decades in the context of phase transformation and plasticity \cite{seeger}.
In another case, however, with an exception of a finite set of particles, all particles till $\pm\infty$ are in the same phase so that this configuration can be associated with a dislocation-dipole or kink-antikink pair. 
From the point of view of a continuum limit of Frenkel-Kontorova model, traditionally known as Sine-Gordon equation, the former type of configurations appear as heteroclinic orbits.
In the same limit, the latter type of configurations appear as homoclinic orbits;
in this sense, the present paper deals with equilibria of this type. 
As it is assumed that all particles, except a few localised within a neighbourhood of second phase, are in one phase, it is sufficient that the onsite potential has at least two local minima that are also global minimum. 
The conventional Frenkel-Kontorova model with a periodic onsite potential can be thus replaced by a model with {\em two-well potential}. Sanders \cite{sanders2} and Atkinson and Cabrera \cite{atkinson} presented the Frenkel-Kontorova lattice model for a special choice of onsite potential that allowed representation of kink-like equilibria in closed form. 
The choice of onsite potentials in this paper is same as that of Atkinson and Cabrera \cite{atkinson} and Weiner and Sanders \cite{sanders2}.

The simplicity of the chosen framework of Frenkel-Kontorova lattice model allows the construction of a family of dislocation-dipole-like equilibria, with separation between the two dislocations of opposite sign taken as a variable.
These equilibria are connected by a suitable path that is found to exhibit a familiar Peierls relief \cite{Kamimura,Dumitrica} (see also Fig. 15-9 on page 544 of \cite{Hirth}). The foundations of such energy landscape based approach for one dimensional lattice model are formed by the researches of Hobart \cite{hobart1} (see also \cite{trusk1} and \cite{sharma3}). 
This formulation effectively relates to a projection of the energy landscape which is helpful in answering questions dealing with transient as well as steady state motion. 
In the presence of a constant external force applied on each particle of the dislocation-dipole configuration, for a given number of particles in the second phase, a path connecting several possible equilibria is found using the concept of order parameter \cite{hobart1}.
The highlight of the paper is a statement of sequential transition between these equilibria that allows an interpretation in terms of the growth of separation between two dislocations of opposite sign. In this context, an exact solution is provided for the case of a piecewise-quadratic potential with two wells as well as the one with an intermediate spinodal region.

The paper is organised as follows. In first section, the one-dimensional lattice model, motivated by Frenkel-Kontorova model, is formulated and the equations of equilibrium are presented. Second section contains the exact expression of equilibrium configurations for two-quadratic wells. Subsequent section deals with a change in energy for a path connecting two equilibrium configurations, which differ by one or more particles in second phase in the presence of constant force. The energy landscape for a special case when equilibrium configurations differ by one particle in second phase is presented along with the the energy changes for transition between configurations which differ by two particles in second phase. 
In fourth section, the effect of spinodal region is studied within the quadratic well approximation.
Fifth section provides discussion of a cascade of transitions between equilibria such that the number of particles in second phase changes sequentially, i.e., one at a time. 
Some remarks related to the Peierls refief and the effect of finite temperature are given in the final section on discussion.
Four appendices appearing at the end of the paper provide some additional expressions, few derivations, and accessory details of calculations.

\section{Lattice model}
\label{section2}
Let the set of integers $\mathbb{Z}$ be identified with the particles constituting a one-dimensional lattice with lattice constant $\varepsilon$. Let $\tilde{u}_{n}$ denote the displacement of $n$th particle, which is located at position $n\varepsilon$ in the lattice, for each $n\in \mathbb{Z}.$ 
Suppose that the lattice is attached to a rigid foundation with on-site energy density $\tilde{w}$ such that $\tilde{w}(0)=\tilde{w}(a)=0, \tilde{w}'(0)=\tilde{w}'(a)=0,$ and $\tilde{w}''(0)=\tilde{w}''(a)=c>0,$ for some $a>0.$ 
Throughout the paper, the notation $f'$ stands for the derivative of the function $f$ with respect to its argument.
It is assumed that each particle interacts with only its nearest neighbour particles through harmonic forces captured by elastic modulus $E$ so that the discrepancy between $\tilde{u}_{n}$ and the displacements of nearest neighbours; thus, $\tilde{u}_{n-1}$ and $\tilde{u}_{n+1}$ contribute to this interaction. 
Due to the on-site potential, the $n$th particle also experiences a force due to the potential energy $\varepsilon \tilde w(\tilde{u}_n).$
As a representative of external bias in this lattice model, an external force per unit length is also considered and it is assumed to be independent of $n$; suppose that this is denoted by $\tilde\sigma.$ 
The total potential energy as a function of the displacement field $\{\tilde{u}_i\}_{i\in\mathbb{Z}}$ of all particles in the lattice is
\begin{eqn}\tilde{\mathcal{E}}(\{\tilde{u}_i\}_{i\in\mathbb{Z}}){:=}\sum\nolimits_{n\in\mathbb{Z}}\{\frac{1}{2}E\varepsilon((\tilde{u}_{n+1}-\tilde{u}_{n})/{\varepsilon})^2+\varepsilon \tilde w(\tilde{u}_{n})-\varepsilon \tilde\sigma\tilde{u}_n\}.
\label{free_energy1}
\end{eqn}
Above leads to the equation of equilibrium:
$\frac{E}{\varepsilon}
(\tilde{u}_{n+1}-2\tilde{u}_{n}+\tilde{u}_{n-1})-\varepsilon \lbrack \tilde w^{\prime}(\tilde{u}_{n})-\tilde\sigma ]=0, \forall n\in\mathbb{Z}.$
In this paper, the definitions are emphasized by ${:=}$ symbol in place of equal sign.
In order to reduce the number of physical parameters, let 
\begin{eqn}
u_{n}{:=}(2\tilde u_{n}/a-1), \sigma{:=}\tilde\sigma /(ac)\text{ and }w(u_n){:=}2/(ca^2)\tilde w(\tilde u_n).
\end{eqn}
Let $u_{n}$ for $n\in\mathbb{Z}$ represent the displacement at lattice site $n$.
Then the equation of equilibrium, in above dimensionless formulation of the lattice model, can be rewritten as
\begin{eqn}
(u_{n+1}-2u_{n}+u_{n-1})-{\kappa}^{2}[w^{\prime}(u_{n})-\sigma ]=0, \forall n\in\mathbb{Z},
\label{discrete model}
\end{eqn}
where\footnote{In this paper, the value $0.5$ of structural constant ${\kappa}$ is often used.}
\begin{eqn}
{\kappa}{:=}\varepsilon \sqrt{2c/{E}}.
\label{discrete_modelkappa}
\end{eqn}
In view of several applications of its expression in the sequel, the potential energy function as a counterpart to \eqref{free_energy1} is stated as
\begin{eqn}
\mathcal{E}(\{u_i\}_{i\in\mathbb{Z}}){:=}\sum\nolimits_{n\in\mathbb{Z}}\{\frac{1}{2}(u_{n+1}-u_{n})^2+{\kappa}^{2}[w(u_{n})-\sigma u_n]\}.
\label{free_energy}
\end{eqn}
\begin{figure}[h]
\begin{center}
\includegraphics[width=.5\linewidth]{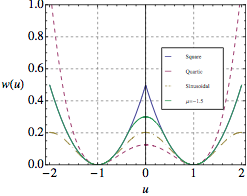}
\end{center}
\caption{\footnotesize 
Double-well onsite potential.}
\label{potential}
\end{figure}
Due to the choice of scaling, the on-site potential $w$ has (global) minima at $\pm 1$ (and also a local maxima at $0$ by continuity). 
For the purpose of obtaining analytical results and closed form expression of desired entities, a special form \cite{atkinson} of $w$ is considered, namely,
\begin{eqn}
w(u){:=}\frac{1}{2}(u+1)^{2}{\Theta}(-u)+\frac{1}{2}(u-1)^{2}{\Theta}(u),
\label{square}
\end{eqn}
where $\Theta$ is the Heaviside function
defined by\footnote{${\Theta}(0)$ does not affect the analysis presented in this paper so it can be left undefined, however, it can be assumed to be $0$ for simplicity.}
\begin{eqn}
{\Theta}(x){:=}1\text{ for }x>0, {\Theta}(x){:=}0, x<0.
\label{heavi}
\end{eqn}
The advantage of such choice of quadratic well potential function $w$ lies in the fact that $w'$ is a piecewise linear function.
The particular $w$ is also shown in Fig. \ref{potential} (as blue curve) alongwith its modified form incorporating an intermediate quadratic region as spinodal region (green curve, with the intermediate curvature $\mu<0$) as well as a quartic well potential function $w(u)=\frac{1}{8}(u-1)^{2}(u+1)^{2}$ (purple curve) and a sinusoidal potential function $w(u)=\frac{1}{\pi^2}(1+\cos\pi x)$ (brown curve) with the restriction that all these functions have the same curvature at $u=\pm1.$ In the context of dislocation, from historical viewpoint, an approximation of a nonlinear function by piecewise linear function has been used by Maradudin \cite{maradudin1}, Sanders \cite{sanders}, Celli and Flytzanis \cite{celli} and Ishioka \cite{ishioka} for a screw dislocation, and by Kratochvil and Indenbom \cite{krato_inden}, Weiner and Sanders \cite{sanders2}, and Atkinson and Cabrera \cite{atkinson} for Frenkel-Kontorova model.

\section{Family of equilibria}
\label{statics}
In the case of the quadratic well potential \eqref{square}, the difference equation \eqref{discrete model} becomes a piecewise-linear difference equation that describes the equilibrium configurations of assumed lattice model,
namely, 
\begin{eqn}
(u_{n+1}-2u_{n}+u_{n-1})-{\kappa}^{2}[u_{n}-{\sigma+1}-2{\Theta}(u_{n})]=0, \forall n\in\mathbb{Z}
\label{disc_eqn}
\end{eqn}
with the boundary conditions 
\[
\lim_{n\rightarrow+\infty}u_n=\lim_{n\rightarrow-\infty}u_n=\text{const}. 
\]

\begin{figure}[h]
\begin{center}
\includegraphics[width=.5\linewidth]{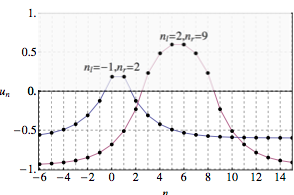}
\end{center}
\caption{\footnotesize 
Equilibria (local minima) for ${\kappa}=0.5$ with $2$ ($\sigma=0.4$) and $6$ ($\sigma=0.05$) particles in the second phase (gray portion).}
\label{statsol}
\end{figure}

Assuming 
\begin{eqn}
n_l<n_r,\text{ with }n_l, n_r\in\mathbb{Z}\text{ such that }u_n>0, \forall n\in(n_l, n_r)\cap\mathbb{Z},\\
\text{ and }
u_n<0, \forall n\notin(n_l, n_r)\cap\mathbb{Z},
\label{solcondns}
\end{eqn}
the equation \eqref{disc_eqn} becomes a system of coupled linear equations and these can be solved using well known methods for linear difference equations \cite{levy}. 
Eventually, a family of 
dislocation-dipole-like stable equilibrium configurations can be expressed as
\begin{eqn}
u_{n}=-1+{\sigma}+\left\{ 
\begin{array}{cc}
A_{n_l-n_r}{\eta}^{n-n_l}, & n\leq n_{l} \\ 
2+B({\eta}^{n-n_r}+{\eta}^{n_l-n}), & n_{l}<n<n_{r} \\ 
A_{n_l-n_r}{{\eta}^{n_r-n}}, & n_{r}\leq n
\end{array}
\right.
\label{sol_D}
\end{eqn}
\begin{eqn}
\text{with }
{\eta} {:=}{\eta}({\kappa})=1+\frac{{\kappa}^{2}}{2}+\frac{{\kappa}}{2}\sqrt{{\kappa}^{2}+4}
\label{def_rho}
\end{eqn}
\begin{eqn}
\text{and }
A_m{:=} A_m({\eta})=2(1-{\eta}^{m+1})/({\eta}+1), B{{:=}}B({\eta})=-2{\eta}/({\eta}+1).
\end{eqn}

An example of the equilibrium configurations given by (\ref{sol_D}) is shown in Fig. \ref{statsol}. 
Due to the form of the chosen on-site potential \eqref{square} and the assumptions \eqref{solcondns}, the equilibria \eqref{sol_D} exist for a specific range of values of $\sigma$. In fact, in \eqref{sol_D} $u_{n_r}\leq0\Rightarrow\sigma\leq\sigma_{upper}$ and $u_{n_r-1}\geq0\Rightarrow\sigma\geq\sigma_{lower}$ with
$\sigma_{upper}=({\eta}-1+2{\eta}^{-n_r+n_l+1})/({\eta}+1), \sigma_{lower}=(1-{\eta}+2{\eta}^{-n_r+n_l+2})/({\eta}+1).$
In other words, the formal solution prescribed by \eqref{sol_D}, with $N=n_r-n_l-1$ particles in the second phase, is {\em admissible} when 
\begin{eqn}
\sigma\in[\sigma_{lower}, \sigma_{upper}]
\label{rangesigUL}
\end{eqn}
\begin{eqn}
\text{where }
\sigma_{upper}{{:=}}({\eta}-1+2{\eta}^{-N})/({\eta}+1)\text{ and }\sigma_{lower}{{:=}}(1-{\eta}+2{\eta}^{-N+1})/({\eta}+1).
\label{def_sigUL}
\end{eqn}
Here, the phrase `$n$th particle is in the first (resp. second) phase', it is meant that $u_n<0$ (resp. $>0$). Thus, all particles with $n\in(n_r, n_l)\cap\mathbb{Z}$ are in second phase and remaining are in first phase according to \eqref{sol_D} provided \eqref{rangesigUL} holds. 

\begin{sidenote}
In the case of single particle in the second phase, say at $n=0$ with $n_l=-1, n_r=1$, the expression corresponding to \eqref{sol_D} is given by
$u_{n}=\sigma-1+2{\eta}^{-|n|}({\eta}-1)/({\eta}+1), \forall n\in\mathbb{Z}.$
\label{solsinglepart}
\end{sidenote}

\begin{sidenote}
There are also unstable (saddle point) equilibria as described by \eqref{sol_D1}, \eqref{sol_D2}, and \eqref{sol_D3} in the appendix~\ref{a_sec1} where the particles with $n\in\{n_l\}, n\in\{n_r\},$ and $n\in\{n_l, n_r\}$ are in degenerate spinodal region, respectively; in this paper 
this means that the displacement is zero for the particles at $n\in\{n_l, n_r\}$. All such equilibria \eqref{sol_D}, \eqref{sol_D1}, \eqref{sol_D2}, and \eqref{sol_D3}, coincide when $\sigma=\sigma_{upper}$ or $\sigma=\sigma_{lower}$ for given $n_l, n_r.$ 
\end{sidenote}

Within the simple framework of the case of the quadratic well potential \eqref{square},
when $\sigma$ varies but remains within the upper and lower bounds (\ref{def_sigUL}) for given number of particles in the second phase, the solution profile essentially shifts along the $u$ axis (see Fig. \ref{statsol} for the axes referred) in view of \eqref{sol_D}.
As described in the appendix~\ref{a_sec1}, the above mentioned stable equilibria \eqref{sol_D} are local minima and 
unstable equilibria 
\eqref{sol_D1}, \eqref{sol_D2}, and \eqref{sol_D3} are saddle points of the energy \eqref{free_energy}.
When $\sigma$ equals either $\sigma_{lower}$ or $\sigma_{upper}$ then two particles located at left and right side of the second phase region are in the degenerate spinodal region and the lattice configuration becomes unstable. As soon as $\sigma$ increases beyond these limits the equilibrium configuration transforms into another equilibrium with an increase or decrease in the number of particles in the second phase and the transition continues until a configuration is reached for which upper and lower bounds contain the given value of external force $\sigma,$ if such configuration exists. Indeed there may be many local minima for the same external force and 
a path connecting such configurations by using the concept of order parameter is studied in the remainder of this paper.

\section{Connecting path in energy landscape}
\subsection{Order parameter based framework}
By its nature, near a saddle point equilibrium there are certain directions in the configurational space of lattice where energy decreases and along all other directions energy increases. Thus a reduction of the entire configurational space is possible in the overdamped limit of dynamics so that only a few directions are relevant to describe transition between metastable equilibria. The concept of order parameter describes such reduction. 
Any two equilibrium configurations in the configurational space of the lattice which can be distinguished, based on this notion of order, can be connected by a path defined by the order parameter varying between certain limits.
In the following, the change in energy as a function of an order parameter is studied. 

The equilibrium configuration of lattice is obtained by minimisation of a constrained energy, i.e., the energy \eqref{free_energy} minus a term accounting for order parameter based constraints: 
\begin{eqn}
\mathcal{E}_C(\{u_i\}_{i\in\mathbb{Z}}){{:=}}
\mathcal{E}(\{u_i\}_{i\in\mathbb{Z}})-\sum\nolimits_{i\in\Lambda}\lambda_i u_i,
\label{min_P}
\end{eqn}
\begin{eqn}
\text{with }
u_i={u}^{({\alpha})}_i, \forall i\in\Lambda,
\label{constrain_P}
\end{eqn}
in terms of a parameter ${\alpha}$ described below.
Each Lagrange multiplier $\lambda_i$ is a perturbing force that scouts for the displacement constraint at $i$th particle for each $i\in\Lambda.$ 

Let $\Lambda{:=}\{p_i\}_{i=1}^{\mathtt{N}}$ be a subset of integers containing $\mathtt{N}$ number of positions in the lattice where the displacement is constrained. The equations of equilibrium for each $n\in\mathbb{Z}$ obtained from minimisation of $\mathcal{E}_C$ (\ref{min_P}) are
\begin{eqn}
-(u_{n+1}-2u_{n}+u_{n-1})+{\kappa}^{2}[u_{n}-{\sigma+1}-2{\Theta}(u_{n})]-\sum\nolimits_{i\in\Lambda}\lambda_i\delta_{i, n}=0, \forall n\in\mathbb{Z}
\label{disceqn_P}
\end{eqn}
along with the constraints (\ref{constrain_P}). In (\ref{disceqn_P}), $\delta_{i, n}=1$ if $i=n$ and $0$ otherwise is the Kronecker delta. 
Here, it is convenient to denote a general list of order parameters by ${\upalpha}=({\alpha}_i)_{i\in\Lambda}$ with the number of components equal to $\mathtt{N}.$ 
A transition from one configuration $\{u_n^{(0)}\}_{n\in\mathbb{Z}}$ (with ${\alpha}_i=0$ for all $i$) to another configuration $\{u^{(1)}\}_{n\in\mathbb{Z}}$ (with ${\alpha}_i=1$ for all $i$) is considered so that $n$th particle in the two equilibrium configurations does not change its phase except on the set $\Lambda$ where it is different. The transition is carried out by varying ${u}^{({\alpha})}_{i}$ continuously between $u^{(0)}_{i}$ and $u^{(1)}_{i}$ for each $i\in\Lambda$. 
By using ${\upalpha},$ the constrained value of displacement (\ref{constrain_P}) can be expressed as
\begin{eqn}
{u}^{({\alpha})}_{i}{{:=}}(1-{\alpha}_i)u^{(0)}_{i}+{\alpha}_i u^{(1)}_{i}, \forall i\in\Lambda.
\label{constraintP}
\end{eqn}
Each component of ${\upalpha}$ lies in the interval $[0,1]$ and, therefore, ${\upalpha}$ can be identified with a vector in the unit cube $[0,1]^{\mathtt{N}}\subset\mathbb{R}^{\mathtt{N}}.$ Moreover each vertex of this unit cube lies at, or in proximity of, a stable equilibrium configuration (local minimum of energy) of the lattice which are given by the condition that all $\lambda_i$s are zero (this includes clearly, $[0,\dotsc,0]$ and $[1,\dotsc,1]$). By definition of a stable equilibrium configuration, there is a small neighbourhood surrounding a vertex of the unitcube. Define the maximal of such neighbourhood of $i$th vertex of the unit cube as the set $S_i.$ The boundary of each set $S_i$ is such that those particles which are members of a particular subset of $\Lambda,$ depending on $i,$ lie in degenerate spinodal region.
The equations of constrained equilibrium (\ref{disceqn_P}) can be solved explictly and the displacement field at the sites in $\Lambda$, i.e., $\{u_n(\{\lambda_i,S_i, \}_{i\in\Lambda},{\upalpha})\}_{n\in\mathbb{Z}}$, can be found. Following this $\lambda_{i}=\lambda_{i}(\{{S_i}\}_{i\in\Lambda},{\upalpha})$ can be obtained from the constraints (\ref{constrain_P}) and finally $\{u_n(\{{S_i}\}_{i\in\Lambda},{\upalpha})\}_{n\in\mathbb{Z}}$ can be determined.
The boundaries of the sets ${S_i}$ are determined by the condition $u_{m}(\{{S_i}\}_{i\in\Lambda},{\upalpha})=0$ for each $m\in\Lambda$ and thus, finally, $\{u_n({\upalpha})\}_{n\in\mathbb{Z}}.$

The piecewise-linear difference equations (\ref{disceqn_P}) allow closed form solution that can be written as
\begin{eqn}
u^{({\alpha})}_{n}=u_{n}^{(0)}+\sum\nolimits_{m\in\Lambda}\varpi_m{\alpha}_m{\mathcal{U}}_{n-m}, \forall n\in\mathbb{Z}
\label{ualphasol}
\end{eqn}
\begin{eqn}
\text{where }
{\mathcal{U}}_{n}=2{\eta}^{-|n|}({\eta}-1)/({\eta}+1)\forall n\in\mathbb{Z},
\label{Greensol}
\end{eqn}
and the weights $(\varpi_i)_{i\in\Lambda}$ come from the solution of the linear set of equations 
\begin{eqn}
\sum\nolimits_{m\in\Lambda}\varpi_m{\mathcal{U}}_{n-m}=u_{n}^{(1)}-u_{n}^{(0)}, \forall n\in\Lambda.
\label{condvarpi}
\end{eqn}
Recall that ${\eta}$ is defined by \eqref{def_rho}.
\begin{sidenote}
The expression $\sigma-1+{\mathcal{U}}_n$ is same as the solution (\ref{sol_D}) for $n\le n_l=-1, n\ge n_r=1;$ recall Remark \ref{solsinglepart}.
It is easy to verify that $\frac{1}{2}{\mathcal{U}}_n$ is in fact the Green's function in the sense that it is the solution of the equation
\begin{eqn}
-(u_{n+1}-2u_{n}+u_{n-1})+{\kappa}^{2}u_{n}={\kappa}^{2}\delta_{i, 0}.
\label{Greeneq}
\end{eqn}
\end{sidenote}
The Lagrange multipliers, that are forces holding the constraints (\ref{constraintP}), are given by 
$\lambda_m=2\varpi_m({\alpha}_m-{\Theta}(-|{\upalpha}-S_m|))(1-{\eta})^2/{\eta}, \forall m\in\Lambda,$
(using \eqref{def_rho}, it is easy to see that $(1-{\eta})^2/{\eta}=(1+{\eta}^2-2{\eta})/{\eta}={\kappa}^{2}+2-2={\kappa}^{2}$) so that
\begin{eqn}
\lambda_m
=2{\kappa}^2\varpi_m({\alpha}_m-{\Theta}(-|{\upalpha}-S_m|)), \forall m\in\Lambda.
\end{eqn}
Recall \eqref{heavi} for the definition of $\Theta$.
The boundary of $S_m$ is given by \begin{eqn}u_{m}^{(0)}+\sum\nolimits_{n\in\Lambda}\varpi_n{\alpha}_n{\mathcal{U}}_{n-m}=0, \forall m\in\Lambda.\end{eqn}
The expression for Lagrange multipliers can be also rewritten as 
\begin{eqn}
\lambda_m{:=} \lambda_m({\upalpha})=2{\kappa}^{2}\varpi_m({\alpha}_m-{\Theta}(u_{m}^{(0)}+\sum\nolimits_{n\in\Lambda}\varpi_n{\alpha}_n{\mathcal{U}}_{n-m})), \forall m\in\Lambda.
\end{eqn}

The expression (\ref{ualphasol}) describes a constrained path, in the configurational space of the lattice, connecting the two equilibrium configurations $\{u_n^{(0)}\}_{n\in\mathbb{Z}}$ and $\{u_n^{(1)}\}_{n\in\mathbb{Z}}$ with the constraining forces given by $\{\lambda_i\}_{i\in\Lambda}.$ The energy $\mathcal{E}$ of each configuration could be infinite since the lattice contains infinite number of particles. But the change in energy is finite and it is possible to find it along a path connecting two equilibrium configurations. Define $\Psi$ as the change in the energy $\mathcal{E}$ modulo its value at initial configuration and then an expression for $\Psi$ can be derived as shown in the appendix~\ref{a_sec2}. 
In the next section the change in energy associated with the transition from one local minimum to another that includes one more particle in the second phase is presented.

\subsection{Transition from $n$ to $n+1$ particles in second phase}
\begin{figure}[h]
\begin{center}
\includegraphics[width=\linewidth]{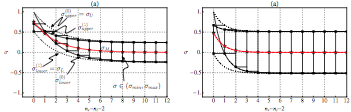}
\end{center}
\caption{\footnotesize 
Effect of ${\kappa}$ over the limits on $\sigma$ for transition between equilibria with a specified number of particles in second phase: (a) ${\kappa}=0.5$, (b) ${\kappa}=1.2$. Star denotes $\sigma_M,$ upper and lower solid box denotes $\sigma_{upper}$ and $\sigma_{lower}$ respectively. The upper and lower line segments of each rectangle denote $\sigma_{max}$ and $\sigma_{min}.$}
\label{range}
\end{figure}

Consider a transition from one equilibrium configuration $\{u_n^{(0)}\}_{n\in\mathbb{Z}}$ with $n_r-n_l-2$ $(>0)$ particles in second phase to another configuration $\{u_n^{(1)}\}_{n\in\mathbb{Z}}$ with $n_r-n_l-1$ particles in the second phase. Choose $\Lambda=\{n_r-1\}$ and let the order parameter be ${\alpha}.$ Therefore, according to \eqref{constraintP}, ${u}^{({\alpha})}_{n_r-1}=(1-{\alpha})u^{(0)}_{n_r-1}+{\alpha} u^{(1)}_{n_r-1}.$ In fact,
$u^{(0)}_{n_r-1}={\sigma-1}+A_{n_l-n_r+1}, u^{(1)}_{n_r-1}={\sigma-1}+2+B({\eta}^{-1}+{\eta}^{n_l-n_r+1}).$ 
Recall that ${\eta}$ is defined by \eqref{def_rho}.
There exists ${\alpha}={\alpha}^{\ast}\in(0,1)$ when the particle located at $n_r-1$ changes its phase from first into second. This critical value of order parameter ${\alpha}^{\ast}$ is given by
\begin{eqn}
{\alpha}^{\ast}{{:=}}\frac{u^{(0)}_{n_r-1}}{(u^{(0)}_{n_r-1}-u^{(1)}_{n_r-1})}=((\sigma-1)({\eta}+1)+2(1-{\eta}^{-n_r+n_l+2}))\frac{1}{2(1-{\eta})}.
\label{def_alphacrit}
\end{eqn}
The dependence of ${\alpha}^{\ast}$ on $\sigma$ and the number of particles in the second phase leads to certain special cases of $\sigma$ and regions of solutions as shown in Fig. \ref{range}.

The energy barrier for trapping regions is maximum when ${\alpha}^{\ast}=0.5$ and the corresponding stress is denoted by $\sigma_M$ given by\footnote{Here $\sigma_M$ is synonymous to the Maxwell stress in the terminology of phase transitions \cite{Ericksen,trusk1}.}
\begin{eqn}
\sigma_M{:=}\frac{2{\eta}^{-n_r+n_l+2}}{({\eta}+1)}.
\label{def_sigM}
\end{eqn}
Thus, \eqref{def_alphacrit} can be expressed as
\begin{eqn}
{\alpha}^{\ast}=\frac{1}{2}+\frac{({\eta}+1)}{2({\eta}-1)}(\sigma_M-\sigma).
\label{def_alphacrit2}
\end{eqn}
The energy barrier for forward transition region 
is maximum when ${\alpha}^{\ast}=1$ and the corresponding stress is 
$\sigma=\sigma_L=-\frac{({\eta}-1)}{({\eta}+1)}+\sigma_M.$
The energy barrier for forward transition is minimum when ${\alpha}^{\ast}=0$ and the corresponding stress is 
$\sigma=\sigma_U=\frac{({\eta}-1)}{({\eta}+1)}+\sigma_M.$
Using \eqref{def_sigUL}, \eqref{def_sigM}, 
since the configuration $u^{(0)}$ contains $n_r-n_l-2$ particles in second phase, it can be observed that $\sigma_{lower}^{(0)}
=(1-{\eta}+2{\eta}^{-n_r+n_l+2+1})/({\eta}+1)
=-({\eta}-1)/({\eta}+1)+{\eta}\sigma_M>\sigma_L$ (as ${\eta}>1$) whereas $\sigma_{upper}^{(0)}=\sigma_U;$
also $\sigma_{lower}^{(1)}
=\sigma_L<\sigma_{lower}^{(0)}$ and $\sigma_{upper}^{(1)}=({\eta}-1+2{\eta}^{-n_r+n_l+1})/({\eta}+1)
=({\eta}-1)/({\eta}+1)+{\eta}^{-1}\sigma_M
<\sigma_U=\sigma_{upper}^{(0)}.$
The requirement that $\{u_n^{(0)}\}_{n\in\mathbb{Z}}$ and $\{u_n^{(1)}\}_{n\in\mathbb{Z}}$ are equilibria at the {\em same} external force $\sigma$ implies that the transition may occur only for 
\begin{eqn}
\sigma\in(\sigma_{min}, \sigma_{max})=(\sigma_{lower}^{(0)},\sigma_{upper}^{(1)}),
\label{def_sigmaxminrange}
\end{eqn}
with
\begin{eqn}
\sigma_{min}=\text{max}\{\sigma_{lower}^{(0)}, \sigma_{lower}^{(1)}\}=\sigma_{lower}^{(0)}\text{ and }\sigma_{max}=\text{min}\{\sigma_{upper}^{(0)}, \sigma_{upper}^{(1)}\}=\sigma_{upper}^{(1)}.
\label{def_sigmaxmin}
\end{eqn}
Note that the interval $[\sigma_L, \sigma_U]$ may not necessarily equal the range of $\sigma$ for admissible transitions as $\sigma_{lower}^{(0)}>\sigma_L$ and $\sigma_{upper}^{(1)}<\sigma_U$. 

\begin{figure}[h]
\begin{center}
\includegraphics[width=.4\linewidth]{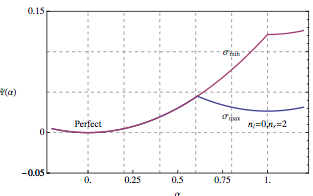}
\end{center}
\caption{\footnotesize 
Example of a Peierls Landscape with minimum and maximum values of $\sigma$ for ${\kappa}=0.5$ and transition from a configuration with $0$ particle in second phase to a configuration with $1$ particle in second phase. Here $\sigma_{\max}$ and $\sigma_{\min}$ correspond to the square dots shown at $n_r-n_l-2=0$ value of the horizontal axis of Fig. \ref{range}.
}
\label{OneDz}
\end{figure}

\begin{sidenote}
For the special case of nucleation of two dislocations with opposite sign, (as shown in Fig. \ref{OneDz}), 
there is one particle in the second phase (recall Remark \ref{solsinglepart}) in the final equilibrium configuration $\{u_n^{(1)}\}_{n\in\mathbb{Z}}$ while the initial configuration $\{u_n^{(0)}\}_{n\in\mathbb{Z}}$ is the single phase (perfect) lattice, also addressed as homogenous state, with $u_n^{(0)}=-1+\sigma, \forall n\in\mathbb{Z}.$ The critical value of order parameter is
${\alpha}^{\ast}=\frac{1}{2}(1-\sigma){({\eta}+1)}/{({\eta}-1)}.$
Using \eqref{def_alphacrit} also, it is found that same expression holds with $n_l=-1, n_r=1$.
Thus in the absence of thermal agitation, the external force required to nucleate two dislocations with opposite sign is given by $\sigma=1.$ 
Therefore, $[\sigma_{min}, \sigma_{max}]=[\sigma_{lower}^{(1)}, \sigma_{upper}^{(1)}]$.
For ${\alpha}^{\ast}=1$ it is found that
$\sigma_L=\sigma_{lower}^{(1)}=1-{2({\eta}-1)}/{({\eta}+1)}$ and for ${\alpha}^{\ast}=0.5$, it is easy to see that 
$\sigma_M={2}/{({\eta}+1)},$ which may or may not lie inside the admissible range of $\sigma$.
On the other hand, $\sigma_{upper}^{(1)}=({\eta}-1+2{\eta}^{-1})/({\eta}+1)<1.$
In particular, $\sigma_{upper}^{(1)}-\sigma_{lower}^{(1)}
=2{\kappa}^2/({\eta}+1)$ and there is a non-trivial barrier at all admissible $\sigma$.
At finite temperature, the energy barrier for may be overcome at any stress $\sigma\in[\sigma_{min}, \sigma_{max}]$ and the dislocations may be nucleated in the lattice. After this nucleation, depending on $\sigma$ and persistence of certain minimum temperature, the two fronts may begin to separate from each other leading to plastic slip of the lattice.
\label{signucl}
\end{sidenote}

\begin{figure}[h]
\begin{center}
\includegraphics[width=.7\linewidth]{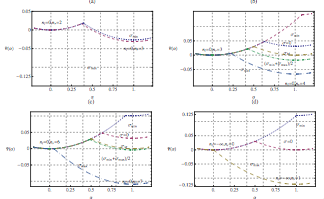}
\end{center}
\caption{\footnotesize 
Example of a Peierls Landscape with different values of $\sigma$ for ${\kappa}=0.5$. 
(a) transition from a configuration with $1$ particle in second phase to a configuration with $2$ particles in second phase, (b) transition from $2$ particles in second phase to $3$, (c) transition from $5$ particles in second phase to $6$, (d) energy landscape for a single dislocation considered as a limiting case of two dislocations of opposite sign far away from each other.
In (a), $\sigma_{\max}$ and $\sigma_{\min}$ correspond to the square dots shown at $n_r-n_l-2=1$ value of the horizontal axis of Fig. \ref{range}; similar scenario holds for other three parts.}
\label{OneD}
\end{figure}

\begin{sidenote}
For the special case of one dislocation (as shown in Fig. \ref{OneD}(d)), $n_r$ is finite and $n_l\rightarrow-\infty$ which gives 
$\overset{\infty}{{\alpha}}^{\ast}=(-\sigma\frac{({\eta}+1)}{({\eta}-1)}+1)\frac{1}{2}.$
In order to initiate the motion of a dislocation the Peierls stress required is $\overset{\infty}{\sigma}_P=({\eta}-1)/({\eta}+1)$ which agrees with the results of Atkinson and Cabrera \cite{atkinson} 
(see also \cite{kresse,sharma14,sharma24}). In this scenario, the maximal energy barrier for both forward and backward transition exists with $\sigma_M=0$ (according to \eqref{def_sigM}) and this is because the configuration with one more or less particle in the second phase is identical to the previous one.
\end{sidenote}

The change in energy, in one-dimensional case, is given by a very simple expression
\begin{eqn}
\Psi({\alpha})=2{\kappa}^2{\mathcal{U}}_0(\frac{1}{2}\varpi^2{\alpha}^2+\varpi({\alpha}^{\ast}-{\alpha}){\Theta}({\alpha}-{\alpha}^{\ast})),
\label{oneD_psi}
\end{eqn}
where, in view of \eqref{condvarpi}, $\varpi$ is given by 
$\varpi=({{u}^{(1)}_{n_r-1}-{u}^{(0)}_{n_r-1}})/{{\mathcal{U}}_0},$
and ${\alpha}^{\ast}$ is given by \eqref{def_alphacrit}.
Recall that the definition of $\Theta$ is given by \eqref{heavi}.
\begin{sidenote}
Using \eqref{sol_D} (also recall the statements preceding \eqref{def_alphacrit}), it is easy to simplify the above expression so that
$\varpi=1.$
\label{varpivalue}
\end{sidenote}
The dependence of $\Psi$ on the number of particles in the second phase is only through dependence on ${\alpha}^{\ast}.$ In Fig. \ref{OneD}(a,b,c) the change in energy for the transition from a configuration involving $1,2,4$ particles in the second phase to $2,3,5$ particles, respectively, is shown. 
The special values of $\sigma$ listed in the figure may be easily computed using the expressions presented in the previous paragraphs in equations \eqref{def_sigUL} and \eqref{def_sigM}.

\begin{sidenote}
It is worth noting that from Fig. \ref{OneD}(a) that there is a very thin range of $\sigma$ which is admissible; in view of Remark \ref{signucl} this is significant as the transition from $1$ to $2$ sites in the second phase appears as a bottleneck. 
\label{trans1to2}
\end{sidenote}

\subsection{Transition from $n$ to $n+2$ particles in second phase}
The transition from one local minimum to another that includes two more particle in the second phase is interesting from the perspective that there is no binding on particles that prohibits simultaneous transition across the phases; in addition to this in some cases there is no intermediate equilibria permitted for a one-one sequential transition to be even possible. The order parameter is a vector of dimension $2.$ 

\begin{figure}[h]
\begin{center}
\includegraphics[width=.6\linewidth]{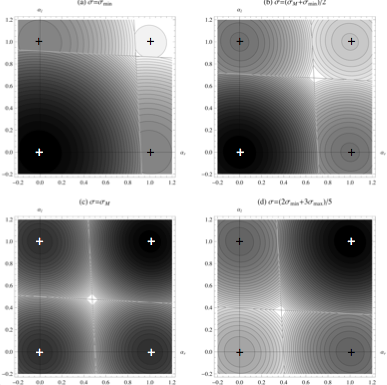}
\end{center}
\caption{\footnotesize 
Two-dimensional energy landscape for ${\kappa}=0.5$ and transition from $n_l=1, n_r=7$ to $n_l=0, n_r=8.$ (a) $\sigma=\sigma_{\min}$ so that at this minimum value of $\sigma$ the two particles in the second phase will not prefer any more particles, (b) $\sigma$ lies between $\sigma_{\min}$ and $\sigma_M$ and in this situation, there is increase in energy or a positive energy barrier associated with particles on either only left or only right side of second phase shift from first phase into second, (c) this is the situation when all energy wells are same except the one with two transitions and this occurs at $\sigma=\sigma_M$, (d) this is the case when the transition to either left or right side is feasible, moreover, when $\sigma>\sigma_{\max}$ the energy barrier for the well located at (0,0) disappears. The lighter sections of the contour plot indicate larger energy while darker sections imply lower. $+$ sign marks the sites with order parameter corresponding to $(0,0)$,$(1,0)$,$(0,1)$ and $(1,1)$.}
\label{twoD}
\end{figure}

Consider a transition from $\{u_n^{(0)}\}_{n\in\mathbb{Z}}$ involving $n_r-n_l-3$ particles in the second phase to $\{u_n^{(1)}\}_{n\in\mathbb{Z}}$ involving $n_r-n_l-1$ particles. Choose $\Lambda=\{n_l+1, n_r-1\}$ and let ${\alpha}_{n_l+1}$ be denoted by ${\alpha}_l$, ${\alpha}_{n_r-1}$ be denoted by ${\alpha}_r$.
Then, according to \eqref{constraintP}, the constraints are given by \begin{eqn}{u}^{({\alpha})}_{n_l+1}=(1-{\alpha}_l)u^{(0)}_{n_l+1}+{\alpha}_l u^{(1)}_{n_l+1}, {u}^{({\alpha})}_{n_r-1}=(1-{\alpha}_l)u^{(0)}_{n_r-1}+{\alpha}_r u^{(1)}_{n_r-1}.\end{eqn}
The solution of the constrained problem is \eqref{ualphasol}, i.e., $u^{({\alpha})}_{n}=u_{n}^{(0)}+\varpi_l{\alpha}_l {\mathcal{U}}_{n-(n_l+1)}+\varpi_r{\alpha}_r {\mathcal{U}}_{n-(n_r-1)}$.
\begin{sidenote}
In this case,
\begin{eqn}
\varpi_{r}&=-\frac{-({u}^{(1)}_{n_r-1}-{u}^{(0)}_{n_r-1}) {\mathcal{U}}_0 + ({u}^{(1)}_{n_l+1}-{u}^{(0)}_{n_l+1}) {\mathcal{U}}_{n_r-n_l-2}}{{\mathcal{U}}_{0}^2 - {\mathcal{U}}_{n_r-n_l-2}^2},\\
\varpi_{l}&=\frac{- ({u}^{(1)}_{n_r-1}-{u}^{(0)}_{n_r-1}) {\mathcal{U}}_{n_r-n_l-2}+({u}^{(1)}_{n_l+1}-{u}^{(0)}_{n_l+1}) {\mathcal{U}}_0 }{{\mathcal{U}}_{0}^2 -{\mathcal{U}}_{n_r-n_l-2}^2}.
\end{eqn}
In view of Remark \ref{varpivalue}, it follows that $\varpi_{r}=\varpi_{l}=1$.
\end{sidenote}

\begin{figure}[h]
\begin{center}
\includegraphics[width=.6\linewidth]{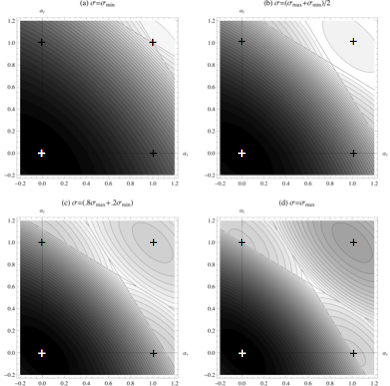}
\end{center}
\caption{\footnotesize 
Two-dimensional energy landscape for ${\kappa}=0.5$ and transition from $n_l=1, n_r=2$ to $n_l=0, n_r=3,$ i.e., from a homogeneous state to a dislocation-dipole.
$+$ sign marks the sites with order parameter corresponding to $(0,0)$,$(1,0)$,$(0,1)$ and $(1,1)$.}
\label{twoDhom}
\end{figure}

In Fig. \ref{twoD}, the contours of the change in the energy for a transition from equilibrium configuration with $5$ particles in second phase to that with $7$ particles is shown and this is for various values of $\sigma$ (relative the transition from $5$ to $6$ particles in second phase) as listed above each contour plot Fig. \ref{twoD}(a, b, c, d). Due to reflection symmetry of the equilibria (\ref{sol_D}) there is a reflection symmetry about the line ${\alpha}_l={\alpha}_r.$ In Fig. \ref{twoD}(c), it is also evident that the energy wells are identical for order parameter $(0, 0)$, $(1, 0)$, and $(0, 1)$ due to $\sigma=\sigma_M.$
There is a saddle point corresponding to either of \eqref{sol_D1}, \eqref{sol_D2}, and \eqref{sol_D3} between any two equilibria corresponding to the local minima of energy landscape. For example see in Fig. \ref{twoD}(b), there is a lowering of energy associated with simultaneous transitions on the left and right side of the second phase region as compared to the increase in energy associated with either only one particle at left side transforming or only the particle at right side changing its phase. This suggests that there is a trapping when the second phase region would not prefer any expansion. For dislocations this would mean that two dislocations with opposite sign may attract each other rather than repel each other. 

\begin{sidenote}
In this simple one dimensional model, the energy barrier for simultaneous change is always greater than that for sequential transition, similar to the case of a phase boundary studied by Sharma and Vainchtein \cite{sharma3}. Following this result, 
a sequential path connecting local minima of energy is described in Appendix \ref{a_sec3}, such that a cascade of transitions occur for given constant force.
\end{sidenote}

In the context of Remark \ref{signucl} and \ref{trans1to2}, it is worth exploring the energy landspace for 2 order parameters starting from the homogeneous state; this is shown in Fig. \ref{twoDhom}. It can be noted in Fig. \ref{twoDhom} from the location of $+$ sign and whether there are any energy well contours around it, that there is an absence of the local minimum besides the initial and final configuration (unlike for example the scenario of Fig. \ref{twoD}) as a consequence of the Remark \ref{trans1to2}.
This is alluded to in the opening sentences of this section.

\section{Equilibria and connecting paths
in a model with 
`spinodal' region}
\label{calcspin}

There are two types of symmetric configuration when the displacement across the peak is equal.
In one case however the particle is at the peak itself and to unravel such configurations it is useful to consider the potential with an intermediate region so that $w$ is a differentiable function.
This brings the motivation to consider a modified form of the expression \eqref{square} so that the potential energy $w$ is given by
\begin{eqn}
w(u){:=}{\frac{1}{2}}\left\{
\begin{array}{cc}
(u+1)^2, & u<-{\upchi}\\
(1-{\upchi})^2+\mu u^2-\mu {\upchi}^2, & \lvert u\rvert\leq {\upchi}\\
(u-1)^2, & u>{\upchi}
\end{array}
\right.,
\label{spinmodel}
\end{eqn}
\begin{eqn}
\text{where }
{\upchi}=\frac{1}{1-\mu}\in(0,1).
\label{spinsize}
\end{eqn}
Alternatively, above is re-written as
\begin{eqn}
w(u)&={\frac{1}{2}} (u+1)^2+\hat{w}(u),\\
&\hat{w}(u){:=}{\frac{1}{2}}\left\{
\begin{array}{cc}
0, & u<-{\upchi}\\
(1-{\upchi})^2+\mu (u^2-{\upchi}^2)-(u+1)^2, & \lvert u\rvert\leq{\upchi}\\
-4u, & u>{\upchi}
\end{array}
\right..
\end{eqn}
Notice that $\mu<0$ and as $\mu\to-\infty$ the spinodal region shrinks so that the model reduces to the two-quadratic wells \eqref{square}. It can be shown that those equilibria when even one particle is present in the spinodal region are not metastable equilibria. In other words, the stable equilibrium configurations can be expressed by \eqref{sol_D} with no particle in the spinodal region. On a careful reading of the conditions \eqref{rangesigUL} on $\sigma$, it is also clear that the admissible range of $\sigma$ also shrinks for finite $\mu$.

Next the transition of configurations similar to that studied in the previous section is presented; in this portion of analysis too, the order parameter is restricted to be first scalar then two dimensional case is considered.

Consider a transition from homogeneous state ${u}^{(0)}$ to that ${u}^{(1)}$ with one atom in the second phase. 
Without loss of generality, it is assumed that the site of the left dislocation and right dislocation is, respectively,
\begin{eqn}
n_l=-\nw-1 , n_r=0,
\label{nlnrh}
\end{eqn}
for the initial state ${u}^{(0)}$.
In this case, thus, there are $\nw$ particles in the second phase of a dislocation-dipole for $\nw>0$.
When $\nw=0$, this corresponds to a homogeneous state.
The stable equilibrium configurations can be expressed as \eqref{sol_D} with \eqref{nlnrh}.
In particular, when $\nw=1$, according to \eqref{sol_D}, it can be considered that
${u}^{(1)}_{-1}=u_{-1}=\sigma+1-{4}/{({\eta}+1)},$
which can be connected via a scalar order parameter starting with the homogenous state ${u}^{(0)}_{n}=-1+\sigma$ (with ${u}^{(0)}_{-1}$ having the same constant value, for example).
Recall that ${\eta}$ is defined by \eqref{def_rho}.
In general, consider the equilibria with $\{{u}^{(0)}_{n}\}$ containing a site with particle at $n=-\nw-1$ in first phase while ${u}^{(1)}_{n}$ having it in the second phase; thus $\Lambda=\{-\nw-1\}$. 
The equilibrium configuration $\{{u}^{({\alpha})}_{n}\}$ of lattice, for such a transition from $\nw-1$ particles in second phase to $\nw$ in second phase, is obtained by minimisation of the energy \eqref{min_P}. It satisfies
\begin{eqn}
-({u}^{({\alpha})}_{n+1}-2{u}^{({\alpha})}_{n}+{u}^{({\alpha})}_{n-1})+{\kappa}^{2}[w'({u}^{({\alpha})}_{n})-{\sigma}]-\sum\nolimits_{i\in\Lambda}\lambda_i\delta_{i, n}=0, \Lambda=\{-\nw-1\}.
\label{min_Peq}
\end{eqn}
In this case, when ${\alpha}_{-\nw-1}={\alpha}_{c1}$, the transition from first phase to spinodal happens and then when ${\alpha}_{-\nw-1}={\alpha}_{c2}$ the transition from spinodal to second phase. For 
${\alpha}_{-\nw-1}\in[0,{\alpha}_{c1}]\cup[{\alpha}_{c2},1]$,
the equation \eqref{disceqn_P} holds with $\Lambda=\{-\nw-1\}$, so that the solution is given by \eqref{ualphasol}, i.e.,
\begin{eqn}
{u}^{({\alpha})}_{n}={u}^{(0)}_{n}+{\alpha}_{-\nw-1}\varpi_{-\nw-1}{\mathcal{U}}_{n+\nw+1}, 
\label{con1dsol}
\end{eqn}
where (in view of \eqref{condvarpi}) 
$\varpi_{-\nw-1}=({{u}^{(1)}_{-\nw-1}-{u}^{(0)}_{-\nw-1}})/{{\mathcal{U}}_0}$
and ${\mathcal{U}}$ is given by \eqref{Greensol},
with 
\begin{eqn}
\lambda_{-\nw-1}=2{\kappa}^{2}\varpi_{-\nw-1}
\left\{
\begin{array}{cc}
{\alpha}_{-\nw-1}, & {\alpha}_{-\nw-1}\in[0,{\alpha}_{c1}]\\
{\alpha}_{-\nw-1}-1, & {\alpha}_{-\nw-1}\in[{\alpha}_{c2},1]
\end{array}
\right..
\end{eqn}
The critical values of ${\alpha}_{-\nw-1}$ are given by
${u}^{(0)}_{-\nw-1}+{\alpha}_{cr1} \varpi_{-\nw-1}{\mathcal{U}}_{0}=-{\upchi}$ and ${u}^{(0)}_{-\nw-1}+{\alpha}_{cr1} \varpi_{-\nw-1}{\mathcal{U}}_{0}=+{\upchi}$,
i.e.,
\begin{eqn}
{\alpha}_{cr1}=\frac{-{\upchi}-{u}^{(0)}_{-\nw-1}}{\varpi_{-\nw-1}{\mathcal{U}}_{0}}, {\alpha}_{cr2}=\frac{+{\upchi}-{u}^{(0)}_{-\nw-1}}{\varpi_{-\nw-1}{\mathcal{U}}_{0}}.
\end{eqn}
For
${\alpha}\in[{\alpha}_{c1},{\alpha}_{c2}]$,
\begin{eqn}
-({u}^{({\alpha})}_{n+1}-2{u}^{({\alpha})}_{n}+{u}^{({\alpha})}_{n-1})+{\kappa}^{2}[\mu{u}^{({\alpha})}_{n}\delta_{-\nw-1, n}+(1-\delta_{-\nw-1, n})(1+{u}^{({\alpha})}_{n})-{\sigma}]\\
-\lambda_{-\nw-1}\delta_{-\nw-1, n}=0,
\end{eqn}
with \eqref{constraintP}.
It is found that \eqref{con1dsol} holds again and that (${\alpha}={\alpha}_{-\nw-1}$)
\begin{eqn}
-{\alpha}\varpi_{-\nw-1}({\mathcal{U}}_{n+\nw+1+1}-2{\mathcal{U}}_{n+\nw+1}+{\mathcal{U}}_{n+\nw+1-1})\\
+{\kappa}^{2}[((\mu-1)({u}^{(0)}_{n}+{\alpha}\varpi_{-\nw-1} {\mathcal{U}}_{n+\nw+1})-1)\delta_{-\nw-1, n}+{\alpha}\varpi_{-\nw-1} {\mathcal{U}}_{n+\nw+1}]\\
-\lambda_{-\nw-1}\delta_{-\nw-1, n}=0,
\end{eqn}
which yields
\begin{eqn}
\lambda_{-\nw-1}={\kappa}^{2}((\mu-1){u}^{(0)}_{-\nw-1}-1+{\alpha}\varpi_{-\nw-1}((\mu-1) {\mathcal{U}}_{0}+2)).
\end{eqn}

\begin{figure}[h]
\begin{center}
\includegraphics[width=.7\linewidth]{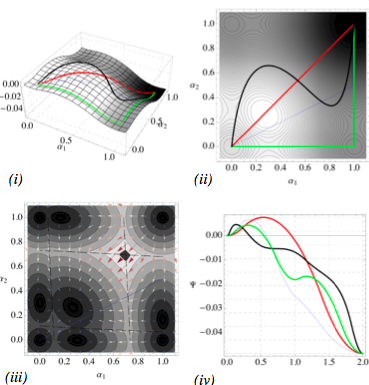}
\end{center}
\caption{\footnotesize 
Two-dimensional energy landscape for $\sigma=0.1$ with ${\kappa}=0.5$ and $\mu=-17/3$ corresponding to the transition from $n_l=1, n_r=7$ to $n_l=0, n_r=8$
(see also Fig. \ref{twoD}(c)).}
\label{figsspin2d1}
\end{figure}

As the work done by the constraint forces along a path in the space of order parameter starting from $0$,
it is found that
\begin{eqn}
\Psi({\alpha})=\int\nolimits_0^{\alpha}\lambda_{-\nw-1}(t)\frac{d}{dt}{u}^{(t)}_{-\nw-1}dt=2{\kappa}^{2}({\frac{1}{2}}{\alpha}^2 \varpi_{-\nw-1}^2{\mathcal{U}}_0+\frac{1}{2}\hat{w}({u}^{({\alpha})}_{-\nw-1})).
\label{Psi1dspin}
\end{eqn}
\begin{sidenote}
If $\Psi'({\alpha})=0$ for ${\alpha}\in(0,1)$, suppose that such critical value of $\Psi$ occurs at ${\alpha}={\alpha}^\ast$, then
${\alpha}^\ast{\mathcal{U}}_0+\frac{1}{2}\hat{w}'({u}^{(0)}_{-\nw-1}+{\alpha}^\ast{\mathcal{U}}_{0}){\mathcal{U}}_0=0,$
i.e.,
\[
{\alpha}^\ast+\frac{1}{2}(\mu u-u-1)|_{u=({u}^{(0)}_{-\nw-1}+{\alpha}^\ast{\mathcal{U}}_{0})}=0,
\]
i.e., ${\alpha}^\ast(1+\frac{1}{2}(\mu-1){\mathcal{U}}_{0})+\frac{1}{2}((\mu-1){u}^{(0)}_{-\nw-1}-1)=0.$
In case of transition from homogeneous state (i.e., $\nw=0$), ${\alpha}^\ast=-\frac{\frac{1}{2}(\mu(\sigma-1)-\sigma)}{1+\frac{1}{2}(\mu-1){\mathcal{U}}_{0}}$ which equals $1$ for $\sigma=\sigma_l=(\mathcal{U}_0-2+\mu(1 - \mathcal{U}_0))/(\mu-1)$ and equals $0$ for $\sigma=\sigma_u=\mu/(\mu-1)=1-{\upchi}$ (as expected); in fact these two limiting values are equal when $\mu=1-2/\mathcal{U}_0=-2/({\eta}-1).$ Thus, for $\mu<-2/({\eta}-1)$, there is a range of $\sigma$, precisely $\sigma\in[\sigma_l, \sigma_u]$, in which the transition from ${u}^{(0)}$ to ${u}^{(1)}$ is possible with a barrier equal to $\Psi({\alpha}^\ast)$.
\end{sidenote}

\begin{figure}[thb!]
\begin{center}
\includegraphics[width=.7\linewidth]{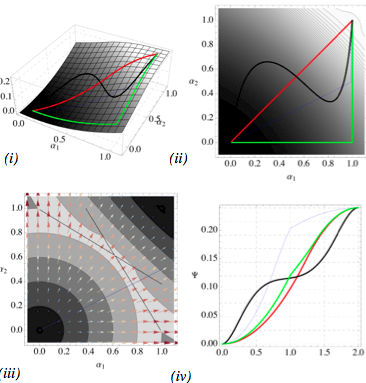}
\end{center}
\caption{\footnotesize 
Two-dimensional energy landscape for $\sigma=0.37$ with ${\kappa}=0.5$ and $\mu=-17/3$ corresponding to the transition from $n_l=1, n_r=2$ to $n_l=0, n_r=3,$ (similar to Fig. \ref{twoDhom}(b)).}
\label{figsspin2d}
\end{figure}

A similar analysis can be carried out for the transition from $n$ to $n+2$ particles in second phase; the relevant details are provided in Appendix \ref{refspinntonp2}.
As a result of the expression \eqref{ualphasol}, by \eqref{min_P}, i.e.,
$\mathcal{E}_C(\{{u}^{({\alpha})}_{n}\})=\mathcal{E}(\{{u}^{({\alpha})}_{n}\})-\sum\nolimits_{i\in\Lambda}\lambda_i {u}^{({\alpha})}_i,$
while $d\mathcal{E}_C(\{{u}^{({\alpha})}_{n}\})/d{u}^{({\alpha})}_{n}=0,$ i.e.,
$\frac{d}{d{u}^{({\alpha})}_{n}}\mathcal{E}(\{{u}^{({\alpha})}_{n}\})=\lambda_i \delta_{i,n},$
the change in the energy is found to be
\begin{eqn}
\Psi({\upalpha})&=\mathcal{E}(\{{u}^{({\alpha})}_{n}\})-\mathcal{E}(\{{u}^{(0)}_{n}\})=\int\nolimits_0^{\upalpha} \frac{d}{dt}\mathcal{E}(\{{u}^{(t)}_{n}\})dt\\
&=\int\nolimits_0^{\upalpha} \frac{d}{d{u}^{({\alpha})}_{n}}\mathcal{E}(\{{u}^{(t)}_{n}\})\frac{d}{dt}{u}^{({\alpha})}_{n}dt=\int\nolimits_0^{\upalpha}\sum\nolimits_n\lambda_n({\alpha})\frac{d}{dt}{u}^{({\alpha})}_{n}dt.
\end{eqn}
As a generalization of \eqref{Psi1dspin}, it is found that above expression yields
\begin{eqn}
\Psi({\upalpha})&=2{\kappa}^{2}\{\frac{1}{2}\sum\nolimits_{m\in\Lambda}\sum\nolimits_{n\in\Lambda}\varpi_n{\alpha}_n{\mathcal{U}}_{n-m}\varpi_m{\alpha}_m+\frac{1}{2}\sum\nolimits_{n\in\Lambda}\hat{w}({u}^{({\alpha}_n)}_{-1})\}.
\end{eqn}
\begin{sidenote}
In the case considered earlier in \S\ref{statics}, $\frac{1}{2}\hat{w}(u)=\frac{1}{4}((u-1)^2-(u+1)^2){\Theta}(u)=-u{\Theta}(u)$ (with $u={u}^{(0)}+{\alpha} \varpi \mathcal{U}_0=-{\alpha}^{\ast} \varpi \mathcal{U}_0+{\alpha} \varpi \mathcal{U}_0=({\alpha} -{\alpha}^{\ast})\varpi \mathcal{U}_0$).
\end{sidenote}
An illustration of the two dimensional order parameter based energy landscape is provided in Fig. \ref{figsspin2d1} for a transition from configuration with $5$ particles in second phase to that with $7$. The figure also reveals the nature of the Lagrange multipliers in this case and also the energy changes along four different kinds of paths in the order parameter space.
Similarly, a transition from homogeneous state to a configuration with $2$ particles in second phase is illustrated in Fig. \ref{figsspin2d}.

Overall, it is clear, from a mathematical viewpoint, that the incorporation of the spinodal region via \eqref{spinsize} leads to smoothening of the Peierls landscape for transition involving configurations that involves one particle crossing the (regular or degenerate) spinodal region in each of the two dislocations (of opposite sign) in the dislocation-dipole.
However, within the confines of the assumed model, the equilibria \eqref{sol_D} remain admissible for the model \eqref{spinmodel} only with small size of spinodal region. As soon as the size \eqref{spinsize} of spinodal region becomes larger the admissible range of $\sigma$ shrinks and the equilibria that exist are not stable any more.

\begin{figure}
\begin{center}
\includegraphics[width=.7\linewidth]{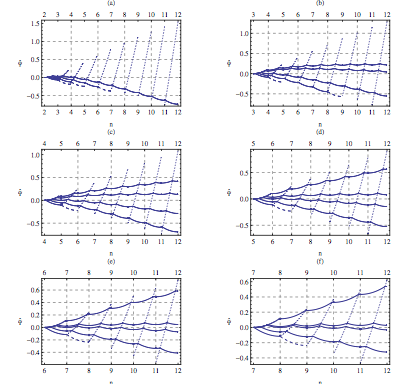}
\end{center}
\caption{\footnotesize 
Example of a Peierls Landscape with various values of $\sigma$ for ${\kappa}=0.5$. The change in energy for a cascade of transitions is shown. In part (a), the change in energy is shown for the cascade of transitions starting with $2$ particle in second phase. In part (b), for $3$ particles as it is clear from the origin of the figures corresponding to $\Psi=0$ and value of $n$ for this, and similarly it can be seen in parts (c), (d), (e) and (f). Solid curve denotes the change in energy, using (\ref{psiall}), for a transition from $n$ particles in second phase to $n+1$ particles where $n$ is the horizontal axis. Blue dashed curve (not grid lines) represents that the Peierls barrier has been crossed indicating the motion of fronts with separation away from one another. Dotted curve denotes the local maxima of the energy barrier between any two stable equilibria. In (b), (c), (d), (e), (f) the top solid curve represents the case $\sigma=\sigma_{\min}$, the second from top curve represents $\sigma=0$, and the third curve is for $\sigma=\sigma_M,$ the fourth curve is drawn for $\sigma=(\sigma_{\max}+\sigma_M)/2$, and the lowest curve is for $\sigma=\sigma_{\max},$ where these values of $\sigma$ are evaluated for the transition from initial configuration with $n$ number of particles in the second phase. In part (a) the curve for $\sigma=0$ is absent because $0<\sigma_M<\sigma_{\min}$ for one particle in second phase.
Here, $\sigma_{\min}$ etc refer to the initial configuration corresponding to the lowest value of $n$ in each plot.
}
\label{peierls}
\end{figure}

\section{Paths incorporating a sequence of transitions}
Let $\{u^{(0)}_n\}_{n\in\mathbb{Z}}$ be the initial configuration with $n_1$ particles in the second phase and the final configuration be $\{u^{(1)}_n\}_{n\in\mathbb{Z}}$ with $n_2$ particles in second phase where $n_2>n_1.$
Consider the quadratic well model {\em without} the spinodal region as discussed in previous section (before its last part). 
Then the change in energy for every single transition can be calculated using the expression (\ref{oneD_psi}). Let ${\alpha}_i$ denote the scalar order parameter for the transition from a configuration with $n_1+i-1$ particles in second phase to a configuration with $n_1+i$ particles and for this transition the change in energy can be expressed as
\begin{eqn}
\Psi({\alpha}_i)=2{\kappa}^2{\mathcal{U}}_0(\frac{1}{2}\varpi_i^2{\alpha}_i^2+\varpi_i({\alpha}_i^*-{\alpha}_i){\Theta}({\alpha}_i-{\alpha}_i^*)),
\label{Psiith}
\end{eqn}
where $\varpi_i$ (for transition at site $n_i$ which lies at the left or right dislocation in the dipole) is
\begin{eqn}
\varpi_i=\frac{{u}^{(1)}_{n_i}-{u}^{(0)}_{n_i}}{{\mathcal{U}}_0},
\label{condvarpisolcas}
\end{eqn}
with (in accordance with \eqref{def_alphacrit}) ${\alpha}_i^*=\{(\sigma-1)({\eta}+1)+2(1-{\eta}^{-n_1-i+1})\}/(2(1-{\eta})), i=1\dotsc N, N=n_2-n_1.$ 
Recall that ${\eta}$ is defined by \eqref{def_rho} and the definition of $\Theta$ is given by \eqref{heavi}.
In view of Remark \ref{varpivalue}, it is noted that $\varpi_i=1$.
Thus the total change $\hat{\Psi}$ in energy for last stage of the transition from the initial order parameter configuration $\{u^{(0)}_n\}_{n\in\mathbb{Z}}$ to $\{u^{(1)}_n\}_{n\in\mathbb{Z}}$ is given by (using \eqref{Psiith})
\begin{eqn}\hat{\Psi}({\alpha}_N){:=}2{\kappa}^2{\mathcal{U}}_0\sum\nolimits_{i=1}^{N-1}(\frac{1}{2}\varpi_i^2{\alpha}_i^2+\varpi_i({\alpha}_i^*-{\alpha}_i){\Theta}({\alpha}_i-{\alpha}_i^*))+\Psi({\alpha}_N)\end{eqn}
If $\sigma$ is such that ${\alpha}_i^*\in[0,1], i=1\dotsc N$ (in other words, $\sigma$ must belong to the intersection of admissible ranges for all such $i$) then using this assumption $\hat{\Psi}({\alpha}_N)$ simplifies to
\begin{eqn}
\hat{\Psi}({\alpha}_N)=2{\kappa}^2{\mathcal{U}}_0\{\sigma(1-{\eta}^2)(N-1)-2{\eta}^{1-n_1}({\eta}^{-N+1}-1)\}/(2(1-{\eta})^2)+\Psi({\alpha}_N)
\label{psiall}
\end{eqn}
An example of this is shown in Fig. \ref{peierls} where the change in energy for a cascade of transitions is illustrated for various values of $\sigma$ for ${\kappa}=0.5$. The details are available in the figure caption.

\begin{sidenote}
As mentioned before the changes in energy of the infinite one-dimensional lattice when the external force $\sigma$ changes cannot be determined. The results presented in Fig. \ref{peierls} show the change in energy as superimposed curves only for convenience. In the situations when $\sigma$ changes, all particles in the lattice are displaced and so mathematically there is infinite increment to the energy \eqref{free_energy}.
\end{sidenote}

\section{Discussion}
\begin{figure}[h]
\begin{center}
\includegraphics[width=.6\linewidth]{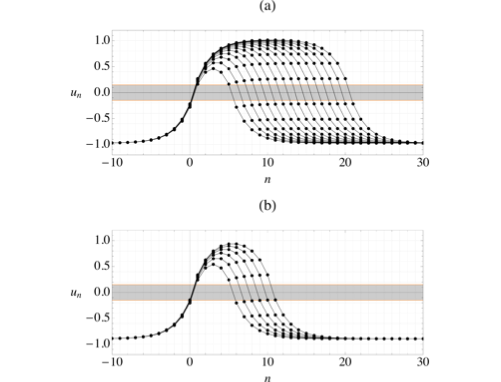}
\end{center}
\caption{\footnotesize 
A family of equilibria (with dipole expanding in the $+n_r$ direction assuming $n_l=0$) beginning with $5$ particles in the second phase for ${\kappa}=0.5$ and (a) $\sigma=\frac{1}{4}\sigma_M,$ (b) $\sigma=\sigma_M$, where $\sigma_M$ corresponds to the transition from $5$ particles in second phase to $6$ (the red curve in Fig. \ref{range}(a) shows $\sigma_M$ dependence).
The gray strip corresponds to the spinodal region for the model discussed in \S\ref{calcspin}.
}
\label{barrier1}
\end{figure}

A visual depiction of the cascade of transitions is also shown in Fig. \ref{barrier1} where a sequence of equilibria \eqref{sol_D} are plotted in (a) and (b) for two values of $\sigma$ respectively as stated in captions.
Upon ignoring the corners or hills in Fig. \ref{peierls}, due to energy barriers caused by lattice trapping in each successive transition, it can be noted that there is an `approximate' curve with a visibly negative curvature and there is one point along the $n$ axis where the slope of this curve becomes zero and beyond this point the slope continues to decrease. Call the point at which the maximum of this `approximate' curve is located as $\ell^a.$ The maximum of $\hat\Psi$ is located at $\ell^c$ and it is near $\ell^a$. As shown below, $\ell^c$ can be found analytically. 
The `approximate' curve connecting all local minima is given by
\begin{eqn}
\Delta\hat\Psi(\ell)=2{\kappa}^2{\mathcal{U}}_0\{\sigma(1-{\eta}^2)(\ell-1)-2{\eta}^{1-n_1}({\eta}^{-\ell+1}-1)\}/(2(1-{\eta})^2).
\label{def_base}
\end{eqn}
By this definition when $\ell=N, \hat\Psi({\alpha}_N)=\Delta\hat\Psi(N)+\Psi({\alpha}_N).$ Recall that ${\eta}$ is defined by \eqref{def_rho}.

\begin{figure}[h]
\begin{center}
\includegraphics[width=\linewidth]{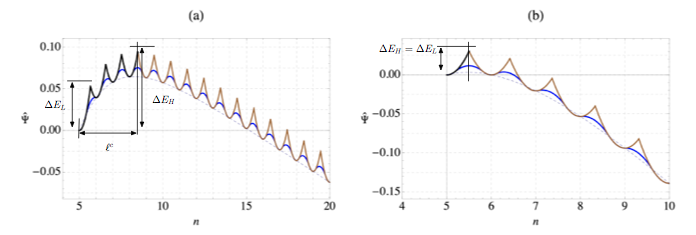}
\end{center}
\caption{\footnotesize 
Energy landspace corresponding to the cascade of transitions of Fig. \ref{barrier1}(a) and (b).
Light solid curve denotes the curve connecting local minima. Thick solid black curve denotes the climb and brown curve denotes sliding. Solid blue curve represents the model having spinodal region with $\mu=-17/3$ (the corresponding spinodal region is shown in Fig. \ref{barrier1} as gray strip). Blue dashed curve denotes the approximate curve \eqref{def_base}.}
\label{barrier}
\end{figure}

The maximum of $\Delta\hat\Psi$ is
\begin{eqn}
\ell^a=-\frac{1}{\ln{\eta}}\ln(\frac{1}{2\ln{\eta}}\sigma({\eta}^2-1){\eta}^{n_1-2}),
\label{def_l}
\end{eqn}
and $\Delta\hat\Psi(\ell^a)={\kappa}^2{\mathcal{U}}_0\{\sigma(1-{\eta}^2)(\ell^a-1+\frac{1}{\ln{\eta}})+2{\eta}^{1-n_1}\}/(1-{\eta})^2.$ The energy barrier for trapping of a dipole is given by
\begin{eqn}
\Delta E_H{:=}\Delta\hat\Psi(\ell^a)+\Psi({\alpha}^*_{\lfloor\ell^a\rfloor}),
\label{def_EH}
\end{eqn}
and clearly, $\ell^c=\ell^a+{\alpha}^*_{\lfloor\ell^a\rfloor},$ where $\lfloor\ell \rfloor$ denotes the greatest integer less than $\ell.$ $ \Delta E_H$ is larger than the energy barrier due to first lattice trapping $\Delta E_L{:=}\Psi({\alpha}_{n_1}^*),$ if $\sigma<\sigma_M$ (for example, see Fig. \ref{barrier} (a)). Also, $\Delta E_H=\Delta E_L$ if $\sigma\ge\sigma_M$. As $\sigma$ increases towards $\sigma_{\max}, \ell^c$ decreases towards the initial number of particles in second configuration, $n_1,$ and indeed $\ell^c={\alpha}_1^*$ when $\sigma=\sigma_M$ (for example, see Fig. \ref{barrier} (b)).

If $\sigma\in(\sigma_{\min}, \sigma_M)$ there is preference towards annihilation and if $\sigma\in(\sigma_M, \sigma_{upper})$ (as previously stated, $\sigma_{upper}=\sigma_{\max}$) there is a preference towards separation of fronts. This is also seen in Fig. \ref{peierls}. For $\sigma\in(\sigma_{\min}, \sigma_M),$ due to thermal excitation, the two fronts can move towards each other, as $\Delta E_H>\Delta E_L,$ and annihilate the dipole. However as soon as $\sigma>\sigma_M, \sigma\in(\sigma_M, \sigma_{upper})$ there is a preference towards a motion that leads to separation of the two fronts away from one another unless during motion there is a coherence between the two fronts as solitary waves described in \cite{sharma4}. This confirms the general principle regarding dislocations in a lattice that two dislocations of opposite sign repel each other and they may attract one another if either the applied stress is too small or the separation between them is small.

In Fig. \ref{barrier}, the solid blue curve refers to the energy profile for the cascade of transitions in the presence of a spinodal region, using the results of \S{calcspin}. From this it is clear that an important role in terms of $\Delta E_L$ is played by the nature of onsite potential model. With an increase in the size of spinodal region, according to \eqref{spinsize} there is a decrease in the value of $|\mu|$ so that the energy barrier naturally reduces. At this point, the last sentences of \S\ref{calcspin} also become highly relevant.

In terms of physical units (for model without the spinodal region), the energy barrier per particle for a transition from $n$ to $n+1$ particles in second phase is \begin{eqn}\tilde\Psi({\alpha}^*)=\frac{E}{4\varepsilon}a^2\Psi({\alpha}^*)=\frac{E}{4\varepsilon}a^2{\kappa}^2{\mathcal{U}}_0\varpi^2{{\alpha}^*}^2\end{eqn}
As $n=n_r-n_l-1\rightarrow\infty,$ $\tilde\Psi({\alpha}^*)\rightarrow\frac{1}{2}\varepsilon{c}\{1-\sigma-2\frac{1}{{\eta}+1}\}^2.$ Using ${\eta}=1+{\kappa}+{\kappa}^2/2+o({\kappa}^2),$ as ${\kappa}\rightarrow0,$ ${\mathcal{U}}_0={\kappa}-{\kappa}^3/4+o({\kappa}^3).$ Choosing the Young's modulus (in three dimensions) $E_0,$ the elastic modulus (in one dimension) is $E\sim E_0a^2, \varpi\sim1$ and suppose $a\sim\varepsilon,$ then
\begin{eqn}
T={\tilde\Psi({\alpha}^*)}/{k_B}=\frac{E_0a^3}{4k_B}\varpi^2{\kappa}^3{{\alpha}^*}^2+o({\kappa}^3){{\alpha}^*}^2.
\end{eqn}
Let $E_0\sim100$ GPa and $a\sim10^{-9}$m, so $T={\tilde\Psi({\alpha}^*)}/{k_B}=\frac{1}{4k_B}{\kappa}^3{{\alpha}^*}^2\times 10^{-16}+o({\kappa}^3){{\alpha}^*}^2$ J per particle. Using $k_B=1.38\times10^{-23}$ J per deg K per particle, $T=1.81\times 10^{6}{\kappa}^3{{\alpha}^*}^2+o({\kappa}^3){{\alpha}^*}^2$ K. With this rough estimate, it can be stated that the energy barriers may be large compared compared to thermal fluctuations at low temperatures. At small $\sigma$ and high temperature, the energy barrier $\Delta E_L$ may be comparable with thermal fluctuations but $\Delta E_H$ may not be overcome by thermal fluctuations along and the external force may need to be increased so that nucleation of dislocation-dipole and propagation of two dislocations is possible. The presence of a relatively small value of the onsite potential elastic constant ${\kappa}$ \eqref{discrete_modelkappa}, in the presence of spinodal region, may lead to a reduction in the energy barrier so it can be overcome even at low temperature.

{\footnotesize{\bf Acknowledgements}:
The partial support of SERB MATRICS grant MTR/2017/000013 is gratefully acknowledged.}

\appendix
\section{
Unstable equilibria}
\label{a_sec1}

\begin{eqn}
u_{n}={\sigma-1}+\left\{ 
\begin{array}{cc}
A_{n_l-n_r}{\eta}^{n-n_l}, & n\leq n_{l} \\ 
2+B^1_{n_l-n_r}{\eta}^{n_l-n}+B^2_{n_l-n_r}{\eta}^{n-n_r}, & n_{l}<n<n_{r} \\ 
-(\sigma-1){{\eta}^{n_r-n}}, & n_{r}\leq n
\end{array}
\right.
\label{sol_D1}
\end{eqn}

\begin{eqn}
u_{n}={\sigma-1}+\left\{ 
\begin{array}{cc}
-(\sigma-1){\eta}^{n-n_l}, & n\leq n_{l} \\ 
2+B^2_{n_l-n_r}{\eta}^{n_l-n}+B^1_{n_l-n_r}{\eta}^{n-n_r}, & n_{l}<n<n_{r} \\ 
A_{n_l-n_r}{{\eta}^{n_r-n}}, & n_{r}\leq n
\end{array}
\right.
\label{sol_D2}
\end{eqn}

\begin{eqn}
u_{n}={\sigma-1}+\left\{ 
\begin{array}{cc}
-(\sigma-1){\eta}^{n-n_l}, & n\leq n_{l} \\ 
2-(\sigma-1)({\eta}^{n-n_r}+{\eta}^{n_l-n})/(1+{\eta}^{n_l-n_r}), & n_{l}<n<n_{r} \\ 
-(\sigma-1){{\eta}^{n_r-n}}, & n_{r}\leq n
\end{array}
\right.
\label{sol_D3}
\end{eqn}
with ${\eta}$ given by \eqref{def_rho}
and $B^1_{n}=-(2{\eta}+{\eta}^n({\eta}-1-({\eta}+1)\sigma))/(({\eta}+1)(1-{\eta}^{2n}))$ and $B^2_n=-(1+{\eta}-2(1+{\eta}^n){\eta}^{1+n}+({\eta}+1)\sigma)/(({\eta}+1)(1-{\eta}^{2n})).$

Consider a small perturbation of any of the equilibria, $\{u^s_i\}_{i\in\mathbb{Z}}$, described by (\ref{sol_D}). Let $u_n=u^s_n+\varepsilon v_n$ with $\varepsilon$ small so that $u_n$ still satisfies the consistency with respect to number of particles in each phase. Then using the fact that $\{u^s_i\}_{i\in\mathbb{Z}}$ is an extrema, it is found that
\begin{eqn}
\mathcal{E}(\{u_i\}_{i\in\mathbb{Z}})-\mathcal{E}(\{u^s_i\}_{i\in\mathbb{Z}})=\varepsilon^2\sum\nolimits_{n\in\mathbb{Z}}\{\frac{1}{2}(v_{n+1}-v_{n})^2+{\kappa}^{2}\frac{1}{2}v^2_{n}\}>0, \notag
\end{eqn}
for all $\varepsilon$ for which the consistency mentioned is {\em not} violated. From the construction it can be seen that such $\varepsilon$ exists for $\sigma\in(\sigma_{lower}, \sigma_{upper})$ for all directions $\{v_i\}_{i\in\mathbb{Z}}$. So equilibria \eqref{sol_D} are local minima. If $\sigma=\sigma_{upper}$ (or $\sigma=\sigma_{lower}$) so that $u^s_N=0$ for some $N$, then one can show that
\begin{eqn}
\mathcal{E}(\{u_i\}_{i\in\mathbb{Z}})&-\mathcal{E}(\{u^s_i\}_{i\in\mathbb{Z}})=\varepsilon^2\sum\nolimits_{n\in\mathbb{Z}}\{\frac{1}{2}(v_{n+1}-v_{n})^2+{\kappa}^{2}\frac{1}{2}v^2_{n}\}\notag\\
&+\varepsilon\sum\nolimits_{n\in\mathbb{Z}}\{(u^s_{n+1}-u^s_n)(v_{n+1}-v_n)+{\kappa}^{2}[(1\pm u^s_n)v_n-\sigma v_n]\}\notag
\end{eqn}
which is positive for all sufficiently small $\varepsilon$ if $v_N=0$ using that $\{u^s_i\}_{i\in\mathbb{Z}}$ is an extrema, but it may be negative for some $\varepsilon$ if $v_N\neq0$ (the directional derivative of $\mathcal{E}$ along $\{v_i\}_{i\in\mathbb{Z}}$ may not exist). For example of the latter, consider $v_n=0, n\neq N, v_N\neq0,$ then $\mathcal{E}(\{u_i\}_{i\in\mathbb{Z}})-\mathcal{E}(\{u^s_i\}_{i\in\mathbb{Z}})<0,$ for $v_N>0$ and $0>\varepsilon>-4{\kappa}^2/{((1+{\kappa}^2)v_N({\eta}+1))}.$ Therefore the equilibria \eqref{sol_D1}, \eqref{sol_D2}, and \eqref{sol_D3} are saddle points.

\section{Derivation of the change in energy}
\label{a_sec2}
The change in the energy is
\begin{eqn}
\Psi({\upalpha})&=\mathcal{E}(\{u_{i}^{({\alpha})}\})-\mathcal{E}(\{u_{i}^{(0)}\})\\
&=\frac{1}{2}\sum\nolimits_n(\sum\nolimits_{m\in\Lambda}\varpi_m{\alpha}_m({\mathcal{U}}_{n+1-m}-{\mathcal{U}}_{n-m}))^2\\&
+\sum\nolimits_n(u_{n+1}^{(0)}-u_{n}^{(0)})(\sum\nolimits_{m\in\Lambda}\varpi_m{\alpha}_m({\mathcal{U}}_{n+1-m}-{\mathcal{U}}_{n-m}))\\&
+\frac{1}{2}{\kappa}^{2}\sum\nolimits_{n\notin\Lambda}[\pm2(1\pm u_n^{(0)})\sum\nolimits_{m\in\Lambda}\varpi_m{\alpha}_m {\mathcal{U}}_{n-m}+(\sum\nolimits_{m\in\Lambda}\varpi_m{\alpha}_m {\mathcal{U}}_{n-m})^2]\\
&-{\kappa}^{2}\sum\nolimits_{n\in\Lambda}\text{ sgn}[u_n^{(0)}+\sum\nolimits_{m\in\Lambda}\varpi_m{\alpha}_m {\mathcal{U}}_{n-m}][u_n^{(0)}+\sum\nolimits_{m\in\Lambda}\varpi_m{\alpha}_m {\mathcal{U}}_{n-m}]\\
&+\sum\nolimits_{n\in\Lambda}\frac{1}{2}{\kappa}^{2}[2u_n^{(0)}\sum\nolimits_{m\in\Lambda}\varpi_m{\alpha}_m {\mathcal{U}}_{n-m}+(\sum\nolimits_{m\in\Lambda}\varpi_m{\alpha}_m {\mathcal{U}}_{n-m})^2-2{u_{n}^{(0)}}]\\
&-\sum\nolimits_n{\kappa}^{2}\sigma(\sum\nolimits_{m\in\Lambda}\varpi_m{\alpha}_m {\mathcal{U}}_{n-m}).
\end{eqn}
Now observe that $\Psi$ must have local extrema at any ${\alpha}_0$ for which all components are either zero or one. So using $\frac{\partial\Psi}{\partial{\alpha}_m}|_{{\alpha}=0}=0,$ obtain
\begin{eqn}
\Psi({\upalpha})&=\frac{1}{2}\sum\nolimits_n\{(\sum\nolimits_{m\in\Lambda}\varpi_m{\alpha}_m({\mathcal{U}}_{n+1-m}-{\mathcal{U}}_{n-m}))^2+{\kappa}^{2}(\sum\nolimits_{m\in\Lambda}\varpi_m{\alpha}_m {\mathcal{U}}_{n-m})^2\}\\
&-2{\kappa}^{2}\sum\nolimits_{n\in\Lambda}\{u_n^{(0)}+\sum\nolimits_{m\in\Lambda}\varpi_m{\alpha}_m {\mathcal{U}}_{n-m}\}{\Theta}[u_n^{(0)}+\sum\nolimits_{m\in\Lambda}\varpi_m{\alpha}_m {\mathcal{U}}_{n-m}],
\end{eqn}
and using $\frac{\partial\Psi}{\partial{\alpha}_m}|_{{\alpha}_m=1, {\alpha}_n=0\forall n\neq m}=0,$
\begin{eqn}
\Psi({\upalpha})&=2{\kappa}^{2}\{\frac{1}{2}\sum\nolimits_{m\in\Lambda}\sum\nolimits_{n\in\Lambda}\varpi_m{\alpha}_n{\mathcal{U}}_{n-m}\varpi_m{\alpha}_m\\
&-\sum\nolimits_{n\in\Lambda}(u_n^{(0)}+\sum\nolimits_{m\in\Lambda}\varpi_m{\alpha}_m {\mathcal{U}}_{n-m}){\Theta}[u_n^{(0)}+\sum\nolimits_{m\in\Lambda}\varpi_m{\alpha}_m {\mathcal{U}}_{n-m}]\}.
\label{genPsi}
\end{eqn}
In one-dimensional case, by explicit calculation also, it can be shown that $\sum\nolimits_n\{({\mathcal{U}}_{n+1}-{\mathcal{U}}_{n})^2+{\kappa}^{2}{\mathcal{U}}_{n}^2\}=2{\kappa}^2{\mathcal{U}}_0.$

\section{Sequential vs simultaneous}
\label{a_sec3}
The energy barrier for sequential transition, where transition occurs effectively through scalar order parameter, is 
\begin{eqn}
\Psi_{sq}({\alpha}^*)=\text{max}\{{\kappa}^2{\mathcal{U}}_0{{\alpha}_l^*}^2, {\kappa}^2{\mathcal{U}}_0({{\alpha}_r^*}^2+2{\alpha}_l^*-1)\},
\end{eqn}
where ${\alpha}_l^*, {\alpha}_r^*$ are given by $u_{n_l+1}^{(0)}+{\alpha}_l^*{\mathcal{U}}_{0}=0, u_{n_r-1}^{(0)}+{\mathcal{U}}_{n_l-n_r+2}+{\alpha}_r^*{\mathcal{U}}_{0}=0.$ It is easy to see that ${\alpha}_r^*<{\alpha}_l^*$ as ${\mathcal{U}}_n>0, \forall n$ and $u_{n_l+1}^{(0)}=u_{n_r-1}^{(0)}<0.$ 
The energy barrier for simultaneous transition is $\Psi_{sm}(\beta^*)={\kappa}^2({\mathcal{U}}_0|{\beta^*}|^2+2{\beta_l^*} {\mathcal{U}}_{n_l-n_r+2}{\beta_r^*}).$ Since $\beta_l=\beta_r$ and ${\mathcal{U}}$ and $u^{(0)}$ have the reflection symmetry, $\Psi_{sm}(\beta^*)=2{\kappa}^2({\mathcal{U}}_0+{\mathcal{U}}_{n_l-n_r+2}){\beta_l^*}^2.$ At $\beta^*,$ $u_{n_l+1}^{(0)}+\beta_l^* ({\mathcal{U}}_{0}+{\mathcal{U}}_{n_l-n_r+2})=0.$ Using $u_{n_l+1}^{(0)}=u_{n_r-1}^{(0)},$
\begin{eqn}
\frac{\Psi_{sm}(\beta^*)}{\Psi_{sq}({\alpha}^*)}=\text{min}\{\frac{2({\mathcal{U}}_0+{\mathcal{U}}_{n_l-n_r+2}){\beta_l^*}^2}{{\mathcal{U}}_0{{\alpha}_l^*}^2}, \frac{2({\mathcal{U}}_0+{\mathcal{U}}_{n_l-n_r+2}){\beta_l^*}^2}{{\mathcal{U}}_0({{\alpha}_r^*}^2+2{\alpha}_l^*-1)}\}.
\end{eqn}
When $0<{\alpha}_l^*<1-{\alpha}_r^*<1,$ it can shown that
\begin{eqn}
\frac{\Psi_{sm}(\beta^*)}{\Psi_{sq}({\alpha}^*)}=2\frac{{\mathcal{U}}_0}{{\mathcal{U}}_0+{\mathcal{U}}_{n_l-n_r+2}}>1,\end{eqn} using ${\mathcal{U}}_0>{\mathcal{U}}_{n_l-n_r+2}>0.$ In the case $1>{\alpha}_l^*>1-{\alpha}_r^*>0, \frac{\Psi_{sm}(\beta^*)}{\Psi_{sq}({\alpha}^*)}>1$ is equivalent to
\begin{eqn}
\frac{1}{{\mathcal{U}}_0+{\mathcal{U}}_{n_l-n_r+2}}>\frac{1}{-u_{n_l+1}^{(0)}}(2+\frac{1}{u_{n_l+1}^{(0)}}({\mathcal{U}}_0+{\mathcal{U}}_{n_l-n_r+2})),
\end{eqn}
or $\beta_l^*+\frac{1}{\beta_l^*}>2,$ but this is an obvious inequality for ${\beta_l^*}\in(0, 1).$

\section{Transition from $n$ to $n+2$ particles in second phase for model with spinodal region}
\label{refspinntonp2}
\subsection{Transition from homogeneous state}
Let
\begin{eqn}
n_l=-h , n_r=-1,
\label{nlnrh2}
\end{eqn}
for the initial state ${u}^{(0)}$.
The stable equilibrium configurations can be expressed as \eqref{sol_D} with \eqref{nlnrh2}.
When $h=2$, 
 it can be considered that
${u}^{(1)}_{-1}={u}^{(1)}_{-2}=u_{-1}=u_{-2}
=\sigma+1-2{\eta}^{-1},$
which can be connected with ${u}^{(0)}_{n}=-1+\sigma$ with ${u}^{(0)}_{-1}$, ${u}^{(0)}_{-2}$ having the same value;
this corresponds to a transition from homogeneous state ${u}^{(0)}$ to that ${u}^{(1)}$ with two atoms in the second phase. 
The equilibrium configuration of lattice, for such a transition from $h-2$ particles in second phase to $h$ in second phase, is obtained by minimisation of the energy
\eqref{min_P}, i.e.,
solving \eqref{min_Peq} with $\Lambda=\{-2,-1\}.$ 
In general, consider $\Lambda=\{-h,-1\},$ and for now $h=2,$ while the case $h>2$ is discussed later.

The case $h=2$ is discussed here for $\Lambda=\{-\nh,-1\},$.
For 
${\upalpha}\in N_{cr1}=\{({\alpha}_{-1},{\alpha}_{-\nh}): {u}^{({\alpha})}_{-1},{u}^{({\alpha})}_{-\nh}<-{\upchi}\}$,
\begin{eqn}
-({u}^{({\alpha})}_{n+1}-2{u}^{({\alpha})}_{n}+{u}^{({\alpha})}_{n-1})+{\kappa}^{2}[1+{u}^{({\alpha})}_{n}-{\sigma}]-\lambda_{-1}\delta_{-1, n}-\lambda_{-\nh}\delta_{-\nh, n}=0.
\label{spineqheq2}
\end{eqn}
For $n\notin\Lambda,$ the solution can be written as
\begin{eqn}
{u}^{({\alpha})}_{n}=-1+{\sigma}+\left\{ 
\begin{array}{cc}
A_{l}{\eta}^{n+\nh+1}, & n\leq-\nh-1 \\ 
A_{r}{{\eta}^{-n}}, & 0\leq n,
\end{array}
\right.
\end{eqn}
such that \eqref{constraintP} holds.
Then, using \eqref{spineqheq2} for $n\in\Lambda$, it is found that
along with ${u}^{({\alpha})}_{-1}=-1+\sigma+A_r{\eta}$ and ${u}^{({\alpha})}_{-\nh}=-1+\sigma+A_l{\eta}$, gives
\begin{eqn}
A_l
=\frac{1}{{\kappa}^{2}}\frac{({\eta}-1)}{({\eta}+1){\eta}}\frac{\lambda_{-1}+\lambda_{-\nh}{\eta}}{{\eta}},\quad
A_r
=\frac{1}{{\kappa}^{2}}\frac{({\eta}-1)}{({\eta}+1){\eta}}\frac{\lambda_{-\nh}+\lambda_{-1}{\eta}}{{\eta}}.
\end{eqn}
It is found that \eqref{ualphasol} holds and
\begin{eqn}
\lambda_{-1}=2{\kappa}^{2}{\alpha}_{-1}\varpi_{-1}, \quad \lambda_{-\nh}=2{\kappa}^{2}{\alpha}_{-\nh}\varpi_{-\nh},
\label{lamh2sol1}
\end{eqn}
where (in view of \eqref{condvarpi})
$\varpi_{-1}=-({-({u}^{(1)}_{-1}-{u}^{(0)}_{-1}) {\mathcal{U}}_0 + ({u}^{(1)}_{-\nh}-{u}^{(0)}_{-\nh}) {\mathcal{U}}_{\nh-1}})/({{\mathcal{U}}_{0}^2 - {\mathcal{U}}_{\nh-1}^2}),$
$\varpi_{-\nh}=({- ({u}^{(1)}_{-1}-{u}^{(0)}_{-1}) {\mathcal{U}}_{\nh-1}+({u}^{(1)}_{-\nh}-{u}^{(0)}_{-\nh}) {\mathcal{U}}_0 })/({{\mathcal{U}}_{0}^2 -{\mathcal{U}}_{\nh-1}^2}).$
The critical value of the components of ${\upalpha}$ is given by ${u}^{(0)}_{-1}+{\alpha}_{-1}\varpi_{-1}{\mathcal{U}}_{-1+1}+{\alpha}_{-\nh}\varpi_{-\nh}{\mathcal{U}}_{-1+\nh}=-{\upchi},$ or ${u}^{(0)}_{-\nh}+{\alpha}_{-1}\varpi_{-1}{\mathcal{U}}_{-\nh+1}+{\alpha}_{-\nh}\varpi_{-\nh}{\mathcal{U}}_{0}=-{\upchi}.$
In general, for
${\upalpha}\in N_{cr1}, N_{cr2}, N_{cr3}, N_{cr4},$
the following relations are obtained,
\begin{eqn}
{\alpha}_{-1}\pm\frac{1}{2}\pm\frac{1}{2}=\frac{1}{2{\kappa}^{2}\varpi_{-1}}\lambda_{-1}, {\alpha}_{-\nh}\pm\frac{1}{2}\pm\frac{1}{2}=\frac{1}{2{\kappa}^{2}\varpi_{-\nh}}\lambda_{-\nh},
\label{lamh2sol}
\end{eqn}
with critical value of ${\alpha}$s given by appropriate conditions.
Certain special points of the unit square are stable equilibria (local minima of energy) of the lattice if and only all $\lambda_i$s are zero (this includes clearly, $[0,0]$ and $[1,1]$).

For
${\upalpha}\notin N_{cr1}\cup N_{cr2}\cup N_{cr3}\cup N_{cr4}=\{({\alpha}_{-1},{\alpha}_{-\nh}): |{u}^{({\alpha})}_{-1}|,|{u}^{({\alpha})}_{-\nh}|>{\upchi}\},$
\begin{eqn}
-({u}^{({\alpha})}_{n+1}-2{u}^{({\alpha})}_{n}+{u}^{({\alpha})}_{n-1})+{\kappa}^{2}[\mu{u}^{({\alpha})}_{n}\delta_{\Lambda, n}+(1-\delta_{\Lambda, n})(1+{u}^{({\alpha})}_{n})-{\sigma}]\\
-\lambda_{-1}\delta_{-1, n}-\lambda_{-\nh}\delta_{-\nh, n}=0.
\end{eqn}
Then
\begin{eqn}
-(-1+\sigma+A_{r}-2{u}^{({\alpha})}_{-1}+{u}^{({\alpha})}_{-\nh})+{\kappa}^{2}[\mu{u}^{({\alpha})}_{-1}-{\sigma}]-\lambda_{-1}=0,
\end{eqn}
\begin{eqn}
-({u}^{({\alpha})}_{-1}-2{u}^{({\alpha})}_{-\nh}-1+\sigma+A_{l})+{\kappa}^{2}[\mu{u}^{({\alpha})}_{-\nh}-{\sigma}]-\lambda_{-\nh}=0,
\end{eqn}
such that \eqref{constraintP} holds.
Then
\begin{eqn}
A_l=\frac{({\eta}-1)^{-1}}{(1+\mu ({\eta}-1)) (3 {\eta}-1+\mu ({\eta}-1)^2 )}(\lambda_{-1} {\eta}+\lambda_{-\nh} (2 {\eta}-1+\mu ({\eta}-1)^2)+A_0),
\end{eqn}
\begin{eqn}
A_r=\frac{({\eta}-1)^{-1}}{(1+\mu ({\eta}-1)) (3 {\eta}-1+\mu ({\eta}-1)^2 )}(\lambda_{-\nh} {\eta}+\lambda_{-1} (2 {\eta}-1+\mu ({\eta}-1)^2)+A_0),
\end{eqn}
\begin{eqn}
\text{where }
A_0={\kappa}^{2}(3 {\eta}-1+\mu ({\eta}-1)^2)(\mu-(\mu-1) \sigma).
\end{eqn}
With $R_1=3 {\eta}-1+\mu ({\eta}-1)^2,$ it is found that $A_0={\kappa}^{2}R_1(\mu-(\mu-1) \sigma),$ while by simplifying $A_l{\eta}^{n+3}$ and $A_r{\eta}^{-n}$, 
it is found that \eqref{ualphasol} holds,
and $\lambda_{-1}$ 
and $\lambda_{-\nh}$ 
are found as
\begin{eqn}
\lambda_{-1}=2\frac{{\kappa}^{2}}{{\eta}({\eta}+1)}(\varpi_{-1}{\alpha}_{-1}{\eta}(2+\mu({\eta}-1))+\varpi_{-\nh}{\alpha}_{-\nh}(\mu-1)({\eta}-1))-{\kappa}^{2}(\mu- (\mu-1) \sigma.
\end{eqn}
\begin{eqn}
\lambda_{-\nh}=2\frac{{\kappa}^{2}}{{\eta}({\eta}+1)}(\varpi_{-1}{\alpha}_{-1}(\mu-1)({\eta}-1)+\varpi_{-\nh}{\alpha}_{-\nh}{\eta}(2+\mu({\eta}-1)))-{\kappa}^{2}(\mu- (\mu-1) \sigma.
\end{eqn}

In a different regime, with ${\upalpha}$ such that it lies in $\{({\alpha}_{-1},{\alpha}_{-\nh}): |{u}^{({\alpha})}_{-\nh}|<{\upchi},|{u}^{({\alpha})}_{-1}|>{\upchi}\}$,
\begin{eqn}
-(-1+\sigma+A_{r}-2{u}^{({\alpha})}_{-1}+{u}^{({\alpha})}_{-\nh})+{\kappa}^{2}[1+{u}^{({\alpha})}_{-1}-{\sigma}]-\lambda_{-1}=0,
\end{eqn}
\begin{eqn}
-({u}^{({\alpha})}_{-1}-2{u}^{({\alpha})}_{-\nh}-1+\sigma+A_{l})+{\kappa}^{2}[\mu{u}^{({\alpha})}_{-\nh}-{\sigma}]-\lambda_{-\nh}=0,
\end{eqn}
such that \eqref{constraintP} holds.
Then
\begin{eqn}
A_l=\frac{({\eta}-1)^{-1}}{(2+\mu ({\eta}-1)){\eta}^2}(\lambda_{-1} {\eta}+\lambda_{-\nh}{\eta}^2+{\kappa}^{2}{\eta}^2(\mu-\sigma(\mu-1)),
\end{eqn}
\begin{eqn}
A_r=\frac{({\eta}-1)^{-1}}{(2+\mu ({\eta}-1)){\eta}^2}(\lambda_{-\nh} {\eta}+\lambda_{-1}(2{\eta}-1+\mu({\eta}-1)^2)+{\kappa}^{2}{\eta}(\mu-\sigma(\mu-1)).
\end{eqn}
Simplifying $A_l{\eta}^{n+3}$ and $A_r{\eta}^{-n}$, 
it is found that \eqref{ualphasol} holds,
and also \eqref{lamh2sol1}${}_1$ holds,
while
\begin{eqn}
\lambda_{-\nh} =-2{\kappa}^{2}\varpi_{-1}{\alpha}_{-1}{\eta}^{-1}\frac{({\eta}-1)(1-\mu)}{{\eta}+1} 
+\frac{(2+\mu ({\eta}-1))}{{\eta}+1}2\varpi_{-\nh}{\alpha}_{-\nh}{\kappa}^{2}
-{\kappa}^{2}(\mu-\sigma(\mu-1)).
\end{eqn}
In another regime, with ${\upalpha}$ such that it lies in $\{({\alpha}_{-1},{\alpha}_{-\nh}): |{u}^{({\alpha})}_{-1}|<{\upchi},|{u}^{({\alpha})}_{-\nh}|>{\upchi}\}$,
after simplifying $A_l{\eta}^{n+3}$ and $A_r{\eta}^{-n}$, it is found that
\eqref{ualphasol} holds, and also it is found that $\lambda_{-\nh}$ is given by \eqref{lamh2sol1}${}_2$ while
\begin{eqn}
\lambda_{-1} =-2{\kappa}^{2}\varpi_{-\nh}{\alpha}_{-\nh}{\eta}^{-1}\frac{({\eta}-1)(1-\mu)}{{\eta}+1} 
+\frac{(2+\mu ({\eta}-1))}{{\eta}+1}2\varpi_{-1}{\alpha}_{-1}{\kappa}^{2}
-(\mu-\sigma(\mu-1)).
\end{eqn}

\subsection{Transition from existing dipole}
The case $h>2$ is discussed here for $\Lambda=\{-h,-1\},$.
For 
${\upalpha}\in N_{cr1}=\{({\alpha}_{-1},{\alpha}_{-h}): {u}^{({\alpha})}_{-1},{u}^{({\alpha})}_{-h}<-{\upchi}\}$, the particles on the left and right sides of the bump are in the first phase while others are in the second phase. For example, the equation of equilibrium is
\begin{eqn}
-({u}^{({\alpha})}_{n+1}-2{u}^{({\alpha})}_{n}+{u}^{({\alpha})}_{n-1})+{\kappa}^{2}[1+{u}^{({\alpha})}_{n}-{\sigma}]-\lambda_{-1}\delta_{-1, n}-\lambda_{-h}\delta_{-h, n}=0,
\label{spineqhgt2}
\end{eqn}
for $n\notin\{-h+1, \dotsc,-2\}.$
For $n\notin\Lambda,$ the solution can be written as
\begin{eqn}
{u}^{({\alpha})}_{n}=-1+{\sigma}+\left\{ 
\begin{array}{cc}
A_{l}{\eta}^{n+h+1}, & n\leq-h-1 \\ 
2+(B_l{\eta}^{n+h+1}+B_r{\eta}^{-n}), &-h<n<-1 \\ 
A_{r}{{\eta}^{-n}}, & 0\leq n,
\end{array}
\right.
\end{eqn}
such that \eqref{constraintP} holds.
Then, using \eqref{spineqhgt2} for $n\in\Lambda$, it is found that
\begin{eqn}
A_l=2\frac{{\eta}^{-1}-{\eta}^{1-h}}{{\eta}+1}+\frac{1}{{\eta}^2-1}(\lambda_{-h}+\lambda_{-1}{\eta}^{1-h}),\\
A_r=2\frac{{\eta}^{-1}-{\eta}^{1-h}}{{\eta}+1}+\frac{1}{{\eta}^2-1}(\lambda_{-1}+\lambda_{-h}{\eta}^{1-h}),
\end{eqn}
\begin{eqn}
\text{and }
B_l=2\frac{1-{\eta}}{{\eta}^2-1}{\eta}^{1-h}+\frac{\lambda_{-1}}{{\eta}^2-1}{\eta}^{1-h},
B_r=2\frac{1-{\eta}}{{\eta}^2-1}{\eta}^{1-h}+\frac{\lambda_{-h}}{{\eta}^2-1}{\eta}^{1-h}.
\end{eqn}
After some simplifications,
it is found that \eqref{lamh2sol1} holds,
where $\varpi_{-1}, \varpi_{-h}$ are given the same expressions as in the context of \eqref{lamh2sol1}, except that $\varpi_{-2}$ is replaced by $\varpi_{-h},$ and \eqref{ualphasol} holds.
The critical value of ${\upalpha}$ is given by either ${u}^{(0)}_{-1}+{\alpha}_{-1}\varpi_{-1}{\mathcal{U}}_{0}+{\alpha}_{-h}\varpi_{-h}{\mathcal{U}}_{-1+h}=-{\upchi},$ or ${u}^{(0)}_{-h}+{\alpha}_{-1}\varpi_{-1}{\mathcal{U}}_{-h+1}+{\alpha}_{-h}\varpi_{-h}{\mathcal{U}}_{-h+h}=-{\upchi}.$
In general, for ${\upalpha}\in N_{cr1}, N_{cr2}, N_{cr3}, N_{cr4},$ such that it lies in $\{({\alpha}_{-1},{\alpha}_{-h}): |{u}^{({\alpha})}_{-1}|,|{u}^{({\alpha})}_{-h}|>{\upchi}\}$,
it is easy to see that \eqref{lamh2sol} holds.
This completes the discussion for four regimes of the order parameter space,
${\upalpha}\in N_{cr1}\cup N_{cr2}\cup N_{cr3}\cup N_{cr4}.$

Certain special points of the unit square are stable equilibria (local minima of energy) of the lattice if and only all $\lambda_i$s are zero (this includes clearly, $[0,0]$ and $[1,1]$).

For
${\upalpha}\notin N_{cr1}\cup N_{cr2}\cup N_{cr3}\cup N_{cr4},$ such that it lies in $\{({\alpha}_{-1},{\alpha}_{-h}): |{u}^{({\alpha})}_{-1}|,|{u}^{({\alpha})}_{-h}|<{\upchi}\}$,
\begin{eqn}
-({u}^{({\alpha})}_{n+1}-2{u}^{({\alpha})}_{n}+{u}^{({\alpha})}_{n-1})+{\kappa}^{2}[\mu{u}^{({\alpha})}_{n}\delta_{\Lambda, n}+(1-\delta_{\Lambda, n})(1+{u}^{({\alpha})}_{n})-{\sigma}]\\
-\lambda_{-1}\delta_{-1, n}-\lambda_{-h}\delta_{-h, n}=0,
\label{spineqhgt3}
\end{eqn}
for $n\notin\{-h+1, \dotsc,-2\}.$
For $n\notin\Lambda,$ the solution can be written as (same as before)
\begin{eqn}
{u}^{({\alpha})}_{n}=-1+{\sigma}+\left\{ 
\begin{array}{cc}
A_{l}{\eta}^{n+h+1}, & n\leq-h-1 \\ 
2+(B_l{\eta}^{n+h+1}+B_r{\eta}^{-n}), &-h<n<-1 \\ 
A_{r}{{\eta}^{-n}}, & 0\leq n,
\end{array}
\right.
\end{eqn}
such that \eqref{constraintP} holds.
Also ${u}^{({\alpha})}_{-1}=-1+\sigma+A_r{\eta}=1+\sigma+B_l{\eta}^{h}+B_r{\eta}$ and ${u}^{({\alpha})}_{-h}=-1+\sigma+A_l{\eta}=1+\sigma+B_l{\eta}+B_r{\eta}^{h}$.
Here, ${\alpha}_n\in[0,1], n\in\Lambda.$
Then, using \eqref{spineqhgt3} for $n\in\Lambda$, 
such that \eqref{constraintP} holds, it is found that
\begin{eqn}
A_l&=\frac{\lambda_{-1} {\eta}^{h+1}({\eta}+1)-\lambda_{-h} ( (\mu-1) ({\eta}-1) {\eta}^2- (2+\mu ({\eta}-1)) {\eta}^{2 h})}{R_2}+A_0,\\
A_r&=\frac{-\lambda_{-1} ( (\mu-1) ({\eta}-1) {\eta}^2- (2+\mu ({\eta}-1)) {\eta}^{2 h})+\lambda_{-h}{\eta}^{h+1}({\eta}+1)}{R_2}+A_0,\\
B_l&=\frac{\lambda_{-1} {\eta}^{h+1}(2+\mu({\eta}-1))-\lambda_{-h}(\mu-1) ({\eta}-1) {\eta}^2}{R_2}+B_0,\\
B_r&=\frac{-\lambda_{-1}(\mu-1) ({\eta}-1) {\eta}^2+\lambda_{-h}{\eta}^{h+1}(2+\mu({\eta}-1))}{R_2}+B_0,
\end{eqn}
\begin{eqn}
\text{where }
R_2&={-(\mu-1)^2 ({\eta}-1)^3 {\eta}^2+(2+\mu ({\eta}-1))^2 ({\eta}-1) {\eta}^{2 h}},\\
A_0&=\frac{{\eta} ({\eta} (-2+\sigma)-\sigma)-\mu ({\eta}-1) ({\eta}+{\eta}^h) (\sigma-1)+{\eta}^h (2+({\eta}-1) \sigma)}{{\eta} ((\mu-1) ({\eta}-1) {\eta}+(2+\mu ({\eta}-1)) {\eta}^h)},\\
B_0&=\frac{-2+({\eta}-1) \sigma-\mu ({\eta}-1) (1+\sigma)}{((\mu-1)({\eta}-1){\eta}+(2+\mu ({\eta}-1)) {\eta}^{h})}.
\end{eqn}
In fact,
the solution can be written as
\begin{eqn}
{u}^{({\alpha})}_{n}={u}^{({\alpha}^{\ast})}_{n}+\left\{ 
\begin{array}{cc}
(A_{l}-A_0){\eta}^{n+h+1}, & n\leq-h-1 \\ 
((B_l-B_0){\eta}^{n+h+1}+(B_r-B_0){\eta}^{-n}), &-h<n<-1 \\ 
(A_{r}-A_0){{\eta}^{-n}}, & 0\leq n,
\end{array}
\right.,
\end{eqn}
where ${\alpha}^{\ast}$ corresponds to the equilibrium with $2$ particles in the unstable spinodal region;
moreover,
\begin{eqn}
{u}^{({\alpha}^{\ast})}_{n}&=-1+\sigma+\left\{ 
\begin{array}{cc}
A_0{\eta}^{n+h+1}, & n\leq-h-1 \\ 
2+B_0({\eta}^{n+h+1}+{\eta}^{-n}), &-h<n<-1 \\ 
A_0{{\eta}^{-n}}, & 0\leq n,
\end{array}
\right.\\
&={u}^{(0)}_{n}+{\alpha}_{-1}\varpi_{-1}\mathcal{U}_0{\eta}^{-|n+1|}+{\alpha}_{-h}\varpi_{-h}\mathcal{U}_0{\eta}^{-|n+h|}
\end{eqn}
so that (it can be easily checked that the equations with terms involving $A_0$ are also satisfied)
\begin{eqn}
{\alpha}_{-1}\varpi_{-1}\mathcal{U}_0
&={\alpha}_{-h}\varpi_{-h}\mathcal{U}_0\\
&=({\eta}-1)\frac{2 (\mu-1) {\eta}^2+ {\eta}^h (2+ ({\eta}+1) \sigma - \mu (\sigma+1+ {\eta} (\sigma-1)))}{(2+ \mu ({\eta}-1)) {\eta}^h ({\eta}+1)+ (-1+ \mu) {\eta} ({\eta}^2-1)},
\end{eqn}
and let such ${\alpha}$s be denoted by
\begin{eqn}
{\alpha}_{\ast}={\alpha}_{-1}={\alpha}_{-h}.
\end{eqn}
Recall \eqref{Greensol}, that is, ${\mathcal{U}}_{n+1}=\mathcal{U}_0{\eta}^{-|n+1|}, {\mathcal{U}}_{n+h}=\mathcal{U}_0{\eta}^{-|n+h|}.$
By simplifying $(A_{l}-A_0){\eta}^{n+h+1}$, $(A_{r}-A_0){\eta}^{-n}$, and $((B_l-B_0){\eta}^{n+h+1}+(B_r-B_0){\eta}^{-n})$, it is found that
\eqref{ualphasol} can be written as ${u}^{({\alpha})}_{n}={u}^{(0)}_{n}+\sum\nolimits_{m\in\Lambda}{\alpha}_{m} \varpi_{m}{\mathcal{U}}_{n-m}={u}^{({\alpha}^{\ast})}_{n}+\sum\nolimits_{m\in\Lambda}({\alpha}_{m}-{\alpha}^{\ast}) \varpi_{m}{\mathcal{U}}_{n-m}.$
These relations lead to
\begin{eqn}
({\alpha}_{-1}-{\alpha}^{\ast})\varpi_{-1}\mathcal{U}_0{\eta}^{-h}R_2&=\lambda_{-1}{\eta}^{h+1}(2+\mu({\eta}-1))-\lambda_{-h}(\mu-1) ({\eta}-1){\eta}^2,\\
({\alpha}_{-h}-{\alpha}^{\ast})\varpi_{-h}\mathcal{U}_0{\eta}^{-h}R_2&=-\lambda_{-1} (\mu-1) ({\eta}-1) {\eta}^2+\lambda_{-h}{\eta}^{h+1}(2+\mu ({\eta}-1)); 
\end{eqn}
indeed, these can be inverted to obtain $\lambda_{-1}$ and $\lambda_{-h}$ in terms of ${\alpha}_{-1}, {\alpha}_{-h}$.

In a different regime, such that it lies in $\{({\alpha}_{-1},{\alpha}_{-h}): |{u}^{({\alpha})}_{-1}|>{\upchi},|{u}^{({\alpha})}_{-h}|<{\upchi}\}$,
 \begin{eqn}
-(-1+\sigma+A_{r}-2{u}^{({\alpha})}_{-1}+{u}^{({\alpha})}_{-2})+{\kappa}^{2}[1+{u}^{({\alpha})}_{-1}-{\sigma}]-\lambda_{-1}=0,\\
-({u}^{({\alpha})}_{-h+1}-2{u}^{({\alpha})}_{-h}-1+\sigma+A_{l})+{\kappa}^{2}[\mu{u}^{({\alpha})}_{-h}-{\sigma}]-\lambda_{-h}=0,
\label{spineqhgt4}
\end{eqn}
such that \eqref{constraintP} holds.
Then
\begin{eqn}
A_l&=\frac{\lambda_{-1} {\eta}^{-h+1}+\lambda_{-h}}{R_3}({\eta}+1)+{A_{l0}},\\
A_r&=\frac{\lambda_{-1}{\eta}^{-1 - 2 h} ((1 - \mu) ({\eta}-1) {\eta}^3 + (2 + \mu ({\eta}-1)) {\eta}^{1 + 2 h})+\lambda_{-h}{\eta}^{1 - h}(1 + {\eta})}{R_3}+{A_{r0}},\\
&- {\eta}^{1 + h} ({\eta}^2-1) (2 - ({\eta}-1) \sigma + \mu ({\eta}-1) (1 + \sigma)))/{R_3},\\
B_l&=\frac{\lambda_{-1} {\eta}^{1 - h}}{{\eta}^2-1}+B_{l0}, \\
B_r&=\frac{-\lambda_{-1}(\mu-1) ({\eta}-1) {\eta}^{2-2h}+\lambda_{-h}{\eta}^{-h+1}({\eta}+1)}{R_3}+{B_{r0}},\\
\end{eqn}
\begin{eqn}
\text{where }R_3&=(2 + \mu ({\eta}-1)) ({\eta}^2-1),\\
A_{l0}&=({\eta}-1) {\eta}^{-1 - h} (-2 {\eta}^2 + {\eta}^h (2 + ({\eta}-1) \sigma + \mu ({\eta}-1)(1-\sigma)))({\eta}+1)/{R_3},\\
A_{r0}&={\eta}^{-1 - 2 h}(2 (\mu-1) ({\eta}-1)^2 {\eta}^3 + 2 (2 + \mu ({\eta}-1)) ({\eta}-1) {\eta}^{2 h} \\
B_{l0}&=\frac{-2({\eta}-1) r^{1 - h}}{{\eta}^2-1},\\
B_{r0}&={\eta}^{-2 h} (2 (\mu-1) ({\eta}-1)^2 {\eta}^2 - {\eta}^h ({\eta}^2-1) (2 - ({\eta}-1) \sigma + \mu ({\eta}-1) (1 + \sigma)))/{R_3}.
\end{eqn}
In fact,
the solution can be written as
\begin{eqn}
{u}^{({\alpha})}_{n}={u}^{({\alpha}^{\ast})}_{n}+\left\{ 
\begin{array}{cc}
(A_{l}-A_{l0}){\eta}^{n+h+1}, & n\leq-h-1 \\ 
((B_l-B_{l0}){\eta}^{n+h+1}+(B_r-B_{r0}){\eta}^{-n}), &-h<n<-1 \\ 
(A_{r}-A_{r0}){{\eta}^{-n}}, & 0\leq n,
\end{array}
\right.,
\end{eqn}
where ${\alpha}^{\ast}$ corresponds to the equilibrium with $1$ particle in the unstable spinodal region.
In fact,
the solution can be written as
\begin{eqn}
{u}^{({\alpha})}_{n}={u}^{({\alpha}^{\ast})}_{n}+\left\{ 
\begin{array}{cc}
(A_{l}-A_{l0}){\eta}^{n+h+1}, & n\leq-h-1 \\ 
((B_l-B_{l0}){\eta}^{n+h+1}+(B_r-B_{r0}){\eta}^{-n}), &-h<n<-1 \\ 
(A_{r}-A_{r0}){{\eta}^{-n}}, & 0\leq n,
\end{array}
\right.,
\end{eqn}
where ${\alpha}^{\ast}$ corresponds to the equilibrium with $2$ particles in the unstable spinodal region. Moreover,
\begin{eqn}
{u}^{({\alpha}^{\ast})}_{n}&=-1+\sigma+\left\{ 
\begin{array}{cc}
A_{l0}{\eta}^{n+h+1}, & n\leq-h-1 \\ 
2+(B_{l0}{\eta}^{n+h+1}+B_{r0}{\eta}^{-n}), &-h<n<-1 \\ 
A_{r0}{{\eta}^{-n}}, & 0\leq n,
\end{array}
\right.\\
&={u}^{(0)}_{n}+{\alpha}_{-1}\varpi_{-1}\mathcal{U}_0{\eta}^{-|n+1|}+{\alpha}_{-h}\varpi_{-h}\mathcal{U}_0{\eta}^{-|n+h|}
\end{eqn}
so that ${\alpha}_{-1}\varpi_{-1}\mathcal{U}_0=0$, and
\begin{eqn}
{\alpha}_{-h}\varpi_{-h}\mathcal{U}_0
&=\frac{({\eta}-1) {\eta}^{-h} (2 (\mu-1) {\eta}^2 + 
 {\eta}^h (2 +({\eta}+1)\sigma - \mu (1 + {\eta} (\sigma-1) + \sigma)))}{(2 + \mu ({\eta}-1)) ({\eta}+1)};
\end{eqn}
let such ${\alpha}$s be denoted by
${\alpha}_{-1}=0, {\alpha}_{\ast}={\alpha}_{-h}.$
It can be easily checked that the equations with terms involving $A_{l0}$ and $A_{r0}$ are also satisfied; indeed,
$A_{l0}-2\frac{{\eta}^{-1}-{\eta}^{{-h+1}}}{{\eta}+1}={\alpha}_{-h}\varpi_{-h}\mathcal{U}_0{\eta}^{-1},$
and
$A_{r0}-2\frac{{\eta}^{-1}-{\eta}^{{-h+1}}}{{\eta}+1}={\alpha}_{-h}\varpi_{-h}\mathcal{U}_0{\eta}^{-h}.$
By simplifying $(A_{l}-A_{l0}){\eta}^{n+h+1}$, $(A_{r}-A_{r0}){\eta}^{-n}$, and $(B_l-B_{l0}){\eta}^{n+h+1}+(B_r-B_{r0}){\eta}^{-n}$, it is found that
\eqref{ualphasol} can be written as ${u}^{({\alpha})}_{n}={u}^{(0)}_{n}+\sum\nolimits_{m\in\Lambda}{\alpha}_{m} \varpi_{m}{\mathcal{U}}_{n-m}={u}^{({\alpha}^{\ast})}_{n}+\sum\nolimits_{m\in\Lambda}({\alpha}_{m}-{\alpha}^{\ast}\delta_{m,-h}) \varpi_{m}{\mathcal{U}}_{n-m}.$
Indeed,
${\alpha}_{-1}\varpi_{-1}\mathcal{U}_0R_3=\lambda_{-1} {\eta}(2 + \mu ({\eta}-1)),$
$({\alpha}_{-h}-{\alpha}^{\ast})\varpi_{-h}\mathcal{U}_0R_3=-\lambda_{-1}(\mu-1) ({\eta}-1) {\eta}^{2-h}+\lambda_{-h}{\eta}({\eta}+1).$
Above relations can be inverted to obtain $\lambda_{-1}$ and $\lambda_{-h}$ in terms of ${\alpha}_{-1}, {\alpha}_{-h}$.
Also similar exercise can be carried out in three other regimes of same type except for combinatorial changes.

\end{document}